\documentclass{aa}
\usepackage{natbib}
\usepackage[varg]{txfonts}
\usepackage{graphicx}
\usepackage{float}
\DeclareMathOperator{\sech}{sech}

\usepackage{color} 

\def\Msun{\ifmmode{\mathrm M_\odot}\else{M$_\odot$}\fi}

\def\Msun{\ifmmode{\mathrm M_\odot}\else{M$_\odot$}\fi}

\def\referee#1{#1}

\def\refereetwo#1{#1}

\def\corr#1{#1}

\begin{document}

\title{Gravitational torques imply molecular gas inflow towards the nucleus of M51}
\author{M.~Querejeta\inst{1}
\and S.~\referee{E.~}Meidt\inst{1}
\and E.~Schinnerer\inst{1}
\and S.~Garc\'{i}a-Burillo\inst{2}
\and C.~\referee{L.~}Dobbs\inst{3}
\and D.~Colombo\inst{4}
\and G.~Dumas\inst{5}
\and A.~Hughes\inst{6,7}
\and C.~Kramer\inst{8}
\and A.~\referee{K.~}Leroy\inst{9}
\and J.~Pety\inst{5,10}
\and K.~\referee{F.~}Schuster\inst{5}
\and T.~\referee{A.~}Thompson\inst{9,11}}

\institute{Max Planck Institute for Astronomy, K\"{o}nigstuhl, 17, 69117 Heidelberg, Germany; \email{querejeta@mpia-hd.mpg.de}
\and Observatorio Astron\'{o}mico Nacional, Alfonso XII, 3, 28014 Madrid, Spain
\and School of Physics and Astronomy, University of Exeter, Stocker Road, Exeter EX4 4QL, UK
\and Department of Physics, 4-181 CCIS, University of Alberta, Edmonton, AB T6G 2E1, Canada
\and Institut de Radioastronomie Millim\'{e}trique, 300 Rue de la Piscine, 38406 Saint Martin d'H\`{e}res, France
\and CNRS, IRAP, 9 Av. Colonel Roche, BP 44346, 31028 Toulouse, France
\and Universit\'{e} de Toulouse, UPS-OMP, IRAP, 31028 Toulouse, France 
\and Instituto Radioastronom\'{i}a Milim\'{e}trica, Avda. Divina Pastora 7, N\'{u}cleo Central, 18012 Granada, Spain
\and Department of Astronomy, The Ohio State University, 140 West 18th Avenue, Columbus, OH 43210, USA
\and Observatoire de Paris, 61 Avenue de l’Observatoire, 75014 Paris, France
\and Center for Cosmology and AstroParticle Physics, The Ohio State University, 191 West Woodruff Avenue, Columbus, OH 43210, USA
}

\date{Received ..... / Accepted .....}

\abstract {The transport of gas towards the centre of galaxies is critical for black hole feeding and, indirectly, it can control active galactic nucleus (AGN) feedback. 
We have quantified the molecular gas inflow in the central $R<1$\,kpc of M51 to be $1\,M_\odot /yr$, using a new gravitational torque map and the molecular gas traced by the \corr{Plateau de Bure interferometer} Arcsecond Whirlpool Survey (PAWS).
 The nuclear stellar bar is responsible for this gas inflow.
We also used torque profiles to estimate the location of dynamical resonances, and the results suggest a corotation for the bar CR$_\mathrm{bar} \sim 20''$, and a corotation for the spiral CR$_\mathrm{sp} \sim 100''$.
We demonstrate how important it is to correct 3.6\,$\mu m$ images for dust emission when gravitational torques are to be computed, and we examine further sources of uncertainty.
Our observational measurement of gas inflow can be compared with nuclear molecular outflow rates and provide useful constraints for numerical simulations.
}


\titlerunning{Molecular Gas Inflow in M51}
\authorrunning{Querejeta et al.}

\maketitle 
\section{Introduction}
\label{Sec:introduction}

The past decades have seen remarkable progress in our understanding of active galactic nuclei (AGN), which are associated with some of the most energetic phenomena in the Universe.
It is now widely accepted that nuclear activity is caused by gas accretion onto a central supermassive black hole \citep[SMBH; e.g.][]{1993ARA&A..31..473A}, and such black holes are expected to exist in all massive galaxies 
\citep{1995ARA&A..33..581K,1998AJ....115.2285M,2000ApJ...539L...9F,2000ApJ...539L..13G,2005SSRv..116..523F,2009ApJ...698..198G,2013ApJ...764..184M}.
 However, fewer than half of the local galaxies that are massive enough to have an SMBH are currently active \citep[$\sim 43\%$ if we include Seyferts, low-ionisation nuclear emission-line
region galaxies, or LINERs, and transition objects; ][]{1997ApJ...487..568H}. It therefore seems natural to associate different levels of nuclear activity with changes in the availability of fuel: either there is an intrinsic dearth of gas in quiescent galaxies, or the transport of the existing gas to the centre is less efficient.

Recent efforts to try and understand gas transport towards the nuclei of nearby galaxies include the project NUclei of GAlaxies \citep[NUGA; e.g.][]{2003A&A...407..485G,2004A&A...414..857C,2005A&A...441.1011G,2007A&A...471..113B,2008A&A...482..133H}.
It is well known that asymmetries in the gravitational potential, such as those caused by bars, drive gas inwards 
\citep{1999MNRAS.304..475M, 2003ASPC..290..411C, 2006LNP...693..143J,2009ApJ...692.1623H};
 simulations including large-scale bars also confirm this
\citep{1987MNRAS.225..653S,1992MNRAS.259..345A,1993ApJ...414..474S,1994ApJ...424...84H,2010MNRAS.407.1529H,2011MNRAS.415.1027H}. Vertical instabilities in the bar that result in boxy- and peanut-shaped bulges can also have a significant impact on the in-plane forces \citep{2015MNRAS.450..229F}.
 Other mechanisms have also been suggested to explain inward motions of gas and AGN feeding, including secondary bars \citep{1989Natur.338...45S}, nuclear spirals \citep{2014A&A...565A..97C}, $m=1$ perturbations, and nuclear warps \citep{2000ApJ...533..850S}. Finally, non-gravitational mechanisms such as viscous torques or dynamical friction of giant molecular clouds against stars have also been invoked to explain gas flows: these can become important near the very centre \citep{1969Natur.223..690L,1972MNRAS.157....1L,2002astro.ph..8113C,2004A&A...414..857C,2005A&A...441.1011G}.

If there is indeed a connection between asymmetries in the gravitational potential and AGN fuelling, and if such an effect is sustained over long enough timescales, an increase of AGN activity in barred galaxies might naturally be expected. However, from an observational perspective, no clear connection between bars and AGN activity has been found so far \citep[e.g.][]{2000ApJ...529...93K,2002ApJ...567...97L,2013ApJ...776...50C}. A possible explanation for this discrepancy would be the existence of some time delay that is still poorly understood (sometimes dubbed the \textit{\textup{timescale conspiracy}}). In particular, it might well be that a hierarchy of mechanisms combine, involving different timescales, which overall conceal the underlying connection \citep{1990Natur.345..679S,2001sac..conf..223C,2003ASPC..290..411C,2004MNRAS.349..270W,2005A&A...441.1011G}. Only recently have observations and simulations begun to achieve the sufficient spatial resolution to track the journey of molecular gas in its last phases of infall towards the central SMBH, showing that a cascade of possibly transient gravitational torques, dynamical friction effects, and filamentary structures can be efficient in transporting the gas to the very centre \citep{2013A&A...558A.124C,2014A&A...565A..97C,2014A&A...567A.125G,2015MNRAS.446.2468E}.

Given the complex interplay of processes suggested so far, it is important to study the feeding of AGN in galaxies with gas imaging at high enough resolution in terms of physical scales, which constrains our targets to nearby galaxies. The grand-design spiral M51 is an ideal target in this sense because of its proximity and the low inclination of the disc \citep[7.6\,Mpc and $i \sim 22^{\circ}$;][]{2002ApJ...577...31C,2014ApJ...784....4C}.
Additionally, as was shown by \citet{2009A&A...496...85G} and \citet{2011A&A...529A..45V}, one of the key points that limit the accuracy in the determination of the radial gas flows is the availability of short-spacing corrections for interferometric molecular gas data. M51 is one of the few galaxies with very high spatial resolution, short-spacing corrected CO(1-0) data 
\citep[the \corr{Plateau de Bure interferometer} Arcsecond Whirlpool Survey, PAWS, 
has a resolution of $1''$ , $\sim 40$\,pc;][]{2013ApJ...779...42S,2013ApJ...779...43P}.
M51 hosts an active galactic nucleus \citep[Seyfert 2,][]{1997ApJS..112..315H}
that is associated with a radio jet \citep{1992AJ....103.1146C}, an outflow in the ionised component \citep{2004ApJ...603..463B}, and also a molecular gas outflow \citep{1998ApJ...493L..63S,2004ApJ...616L..55M,2007A&A...468L..49M,2015ApJ...799...26M}. For the purpose of estimating planar gas flows, we here ignore the molecular gas emission that is not consistent with disc motions.
This emission has a significant \corr{impact} only in the central $\sim 3'' \sim 120$\,pc.

A number of alternatives exist for measuring radial gas flows in nearby galaxies. Based on gas kinematics (traditionally from radio data), the velocity field can be decomposed into its Fourier components, and their radial variations can be used to search for evidence of gas inflow or outflow \citep[e.g.][]{2004ApJ...605..183W}. Also based on gas kinematics, the gas streaming motions can be estimated from analytical solutions for gas orbits in a model that is fitted to the data \citep[e.g.][]{2007A&A...471..113B}. A third option is to obtain the gravitational potential from near-infrared (NIR) images, and weight the implied gravitational torques with the gas distribution to obtain a statistical estimate of gas flows
\citep{1993A&A...274..148G,1995ApJ...441..549Q,2005A&A...441.1011G,2009ApJ...692.1623H,2013ApJ...779...45M}.

In practice, it is hard to extract quantitative measurements from the kinematic method \citep[see][]{2009ApJ...692.1623H}, and modelling tends to rely on too many assumptions and simplifications, which make it only useful for studying some specific effects \citep[e.g.][]{2011A&A...529A..45V}. Additionally, a decomposition of the velocity field becomes increasingly degenerate at low inclinations (the line-of-sight velocities can be difficult to interpret), which is the case for M51.
In our attempt to \textit{\textup{quantify}} gas inflow, we \corr{therefore} opted for the third method.  Another motivation for measuring full torque profiles 
based on accurate stellar mass maps is that it allows us to 
estimate the position of dynamical resonances; most importantly, to identify the number and location of corotation radii in the disc, as we show in Sect.\,\ref{Sec:dynamres}. A complementary, qualitative discussion of the streaming motions in M51, in the line of the first method, and also based on the PAWS dataset, can be found in \citet{2014ApJ...784....4C}.

In the upcoming ALMA and NOEMA era, high-resolution maps of molecular gas will become accessible for an increasing number of nearby galaxies. One of the most important questions to answer is whether the gas that is detected in those galaxies is losing or gaining angular momentum, possibly contributing to the feeding of AGN or nuclear star formation, which can indirectly control possible feedback effects. In this context, it is important to understand the limitations of the observational methods used to calculate gas flows and have the appropriate tools for it, including accurate stellar mass maps, whose importance is emphasised in this paper. Additionally, observational estimates of gas flows play the important role of helping theorists impose constraints on models and simulations of galaxy secular evolution, black hole growth, and AGN feedback.

With the high-resolution map of CO from PAWS and our stellar mass map (based on $3.6\,\mu m$ \textit{Spitzer} IRAC imaging), we are in a position to determine the rates of gas flows in M51 with unprecedented accuracy. Our major goals are 1) to estimate the inflow rate of molecular gas towards the nucleus, 2) use the torque profiles to reasses the location of resonances in M51, and 3) provide an updated study of the uncertainties involved in the observational quantification of gravitational torques and gas flows for nearby galaxies. With these objectives in mind, the paper is organised as follows. After presenting the data in Sect.\,\ref{Sec:data}, we explain the steps involved in the method to calculate torques and gas flows in Sect.\,\ref{Sec:methodology}. The results on torques, inflow, and dynamical resonances are presented in Sect.\,\ref{Sec:results}
 and discussed in Sect.\,\ref{Sec:Discussion}. The dominant sources of uncertainty are examined in Sect.\,\ref{Sec:uncert}, while the details of the tests carried out to quantify them are deferred to Appendix\,\ref{sec:appendix}. We close the paper with a summary and conclusions in Sect.\,\ref{Sec:conclusions}.

\section{Data} 
\label{Sec:data}

Our study relies on NIR imaging from the \textit{Spitzer} Survey of Stellar Structure in Galaxies \citep[S$^4$G;][]{2010PASP..122.1397S} and the molecular gas emission traced by PAWS \citep{2013ApJ...779...42S,2013ApJ...779...43P}. For the purpose of determining the location of dynamical resonances, the information provided by PAWS is complemented with The HI Nearby Galaxy Survey \citep[THINGS;][]{2008AJ....136.2563W} and the HERA CO Line Extragalactic Survey \citep[HERACLES;][]{2009AJ....137.4670L}.

\subsection{NIR data: stellar mass distribution} 
\label{Sec:stellarmass}

NIR emission is often exploited as a stellar mass tracer in nearby galaxies because the light at these wavelengths mainly arises from the old stars that dominate the baryonic mass budget \citep[][]{1993ApJ...418..123R, 1994ApJ...437..162Q}. Additionally, NIR images have the advantage that the effect of extinction is minimised, and biases in the mass-to-light ratio that are due to young O and B stars are also attenuated. For these reasons, the NIR, and, in particular, the first \textit{Spitzer} IRAC band (centred at 3.6\,$\mu m$) has been argued to be an optimal window to trace stellar mass \citep[e.g.][]{2014ApJ...788..144M,2014ApJ...797...55N,2015ApJS..219....5Q}.

However, emission from dust can also contribute significantly to the near-infrared flux, especially locally, as shown by \citet{2012ApJ...744...17M} and \citet{2015ApJS..219....5Q}. Specifically, for M51, the dust emission contributes as much as 34\,\% of the \textup{\textit{\textup{total}}} flux at 3.6\,$\mu m$, and the correction is especially critical in star-forming regions \citep[reaching $\sim 80\%$,][]{2015ApJS..219....5Q}. Therefore, we used the 3.6\,$\mu m$ image of M51 corrected for dust emission using the independent component analysis
 \citep[ICA; ][]{2015ApJS..219....5Q}. After the stellar flux in the 3.6\,$\mu m$ IRAC band is correctly identified, even a single mass-to-light ratio M/L~$ \equiv \Upsilon_{3.6}=0.6$ is applicable with an uncertainty of $\lesssim$0.1\,dex \citep{2014ApJ...788..144M,2014ApJ...797...55N}.

\subsection{CO data: molecular gas distribution} 
\label{Sec:PAWS}

We used the molecular gas amount and distribution probed by the CO(1-0) map of M51 from PAWS 
to perform weighted azimuthal averages of the torques and determine how much gas participates in radial flows.
PAWS offers an exquisite angular resolution of $1''$ ($\sim 40$\,pc) and covers the central 9\,kpc of the galaxy (with uniform azimuthal coverage out to $R=85'' \sim 3$\,kpc). The importance of performing short-spacing corrections to interferometric data for torque-based flow studies has only recently been fully recognised \citep[e.g.][]{2011A&A...529A..45V}. Our PAWS map includes short-spacing corrections based on IRAM 30m single-dish data, which by definition recovers all the flux. We transformed the measured flux into the molecular hydrogen (H$_2$) gas surface density using the Galactic conversion factor $X_\mathrm{CO}=2 \times 10^{20}$\,cm$^{-2}$\,(K\,km\,s$^{-1}$)$^{-1}$
\citep{2013ApJ...779...42S}; the adopted value has no consequence for the measured torque profiles (since the gas map will only serve to weight the azimuthal average of the torques), although it does affect the measured net inflow rate.
For more details on the data reduction, we refer to \citet{2013ApJ...779...43P}.

Within the PAWS field of view, the gas is predominantly molecular, and this becomes even more so towards the centre. Since we are
interested mostly in the central region, where atomic HI gas is largely depleted \citep[HI surface density lower than CO by a factor of 5--10 inside $R<3$\,kpc,][]{2007A&A...461..143S}, and for uniformity reasons, we assumed the gas distribution to be the one traced by the CO observations from PAWS. But we use the information provided by HI (from the THINGS survey, at $6''$ spatial resolution) and by CO at lower resolution 
(HERACLES, $13''$) in Sect.\,\ref{Sec:resonances} to extend the torque analysis to the outer regions of the disc and estimate the positions of resonances.

\section{Measuring gravitational torques and gas inflow rates} 
\label{Sec:methodology}

We focus on the gravitational torques exerted by the stellar potential on the gaseous disc. By definition, a torque is a vectorial quantity, $\pmb{\tau}$, parallel to the axis of the rotation that it would tend to produce: $\pmb{\tau} = \pmb{r} \times \pmb{F}$, the cross product of the position vector and net force acting on a given (test) particle. From this operative definition, the equivalent identity $\pmb{\tau} = d\pmb{L}/dt$ can be derived; therefore, the torque measures the derivative of the angular momentum with respect to time. Since gas needs to lose (gain) angular momentum to move inwards (outwards), torques applied on the gas distribution provide the necessary link to determine gas flows.
Our strategy is to first obtain the gravitational potential using Fourier transforms of the mass distribution (3.6\,$\mu m$ image corrected for dust emission), and, subsequently, the torques implied by that potential are weighted with the molecular gas distribution to determine the gas inflow (outflow) rates as a function of radius.
The code we use is partially based on \texttt{PyPot}  \citep{2009ApJ...692.1623H}. 

\subsection{Deprojection} 
\label{Sec:depro}

Our input is the stellar mass surface density \textit{\textup{projected on the plane of the sky}} (the `stellar mass map' from the ICA method described in Sect.\,\ref{Sec:stellarmass}). Therefore, to obtain the gravitational potential, we first need to deproject this map into the true plane of the galaxy. We did this using the accurate inclination and position angle (PA) kinematic measurements obtained for M51 by \citet{2014ApJ...784....4C}. Their results constrain these values to inclination $i = (22 \pm 5)^\circ$ and $PA = (173 \pm 3)^\circ$. In Sect.\,\ref{Sec:uncert} we examine the effect of deprojecting the image according to the extreme values of $i$ and $PA$ allowed by these uncertainty limits.

Another subtlety that needs to be considered is the fact that a bulge (or any departures from a disc) will become artificially elongated by the deprojection. In the case of M51, \referee{the central structure that has traditionally been called the `bulge' has a disc-like S\'{e}rsic index \citep[$n=0.995$,][]{2015ApJS..219....4S}, and it could in principle have been a pseudo-bulge (or `discy bulge'); however, it has a very similar orientation and extent \citep[$PA = 130^\circ$, $R_\mathrm{e}=16.2''$,][]{2015ApJS..219....4S} as what we will identify as the nuclear bar, suggesting that it is the same stellar structure. Moreover, the low S\'{e}rsic index implies that it is flattened parallel to the plane of the galaxy}. Therefore, provided that the deviation from a plane is very small, the size is modest, and the inclination of M51 is small, it seems justified to ignore the structure of the \referee{nuclear bar} for the deprojection.

\subsection{Obtaining the gravitational potential} 
\label{Sec:pot}

Our technique is based on the idea that the gravitational potential, $\Phi(\pmb{r})$, can be written as the convolution of the mass density, $\rho,$ and the function $1/r$ \citep[e.g.][$\S$ 2.8]{1987gady.book.....B}:

\begin{equation}
\Phi(\pmb{r}) = -G \int \frac{\rho (\pmb{r}')d^3\pmb{r}'}{|\pmb{r}-\pmb{r}'|}.\end{equation}

\noindent
 However, \corr{to} access the true 3D mass density distribution, it is necessary to consider the non-negligible thickness of the galaxy stellar discs. To account for this, we assumed a vertical profile of constant scale-height \citep[in agreement with observations of edge-on galaxies,][]{1989ApJ...337..163W, 1992AJ....103...41B}, which allows us to write

\begin{equation}
\Phi(x,y,z=0) = -G \int \Sigma (x',y') g(x-x',y-y')dx'dy',
\end{equation}

\noindent
where $g(x,y)$ is the modified convolution function, 

\begin{equation}
g(x,y) = \int^{+\infty}_{-\infty} \frac{\rho_z (z)dz}{\sqrt{x^2+y^2+z^2}}.
\end{equation}

In particular, we expect the galaxy to have a vertical distribution similar to an isothermal disc, $\rho_z(z)=\rho_0 \sech^2(z/h)$, with $h \sim 1/12 H_{\mathrm{disc}}$ \citep{1989ApJ...337..163W, 1992AJ....103...41B}. We chose $H_{\mathrm{disc}} = 100''$, based on the Galfit photometric decomposition of M51 performed by \citet{2015ApJS..219....4S}.

An important technical detail is that the convolution is carried out using fast Fourier transforms (FFT) that are computed on a $(2n \times 2n)$ mesh to avoid periodic images \citep{1969JCoPh...4..306H}: the original image occupies one quarter of the grid, and all remaining three quarters are invalid. This means that we initially pad the image with zeroes and eliminate those invalid areas after performing the FFTs.

\begin{figure}[t]
\begin{center}
\includegraphics[trim=0 460 540 0,clip,width=0.5\textwidth]{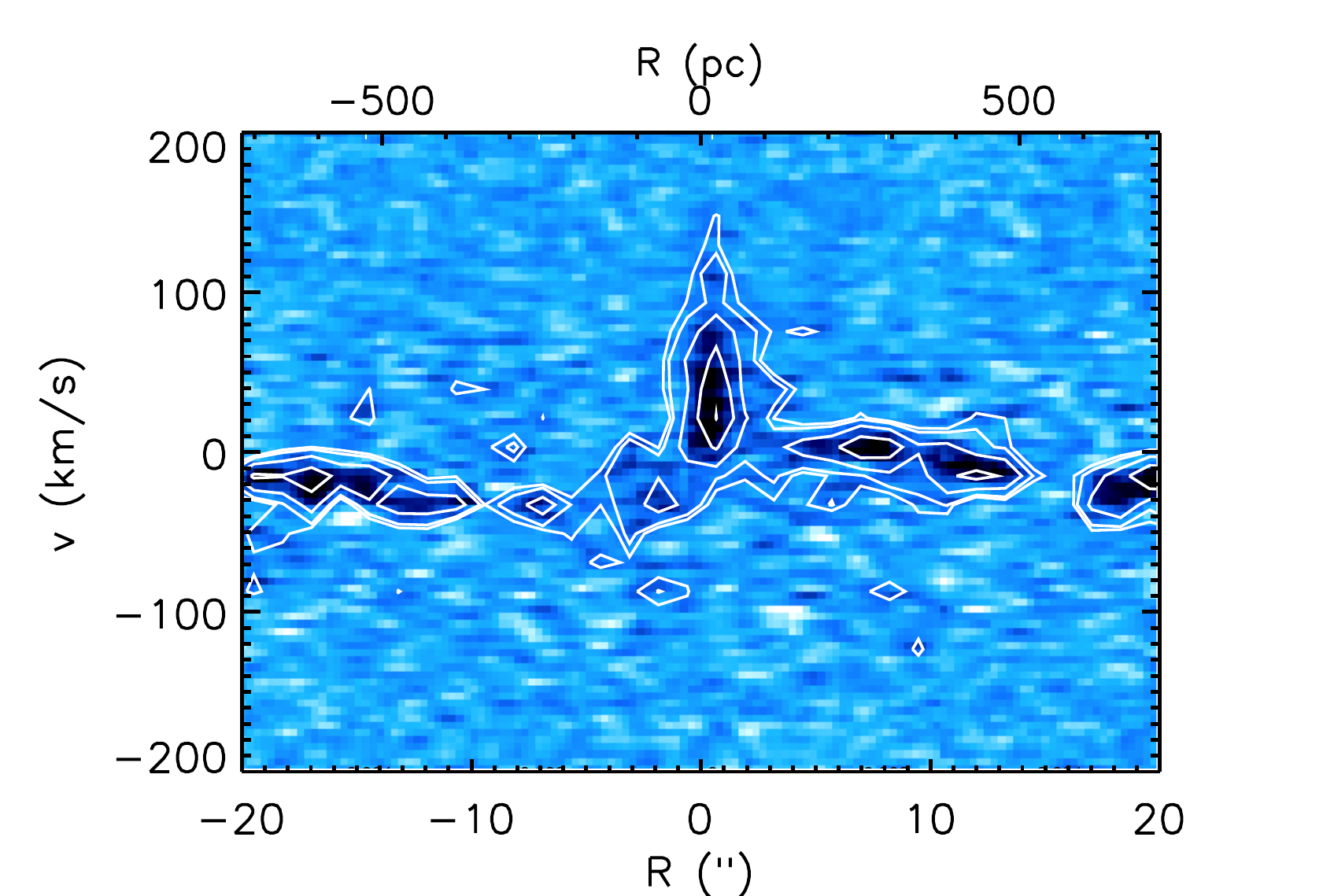}
\end{center}
\caption{Position-velocity diagram of the PAWS $1''$ cube, taken along a position angle of 272$^\circ$ crossing the nucleus. There is highly redshifted CO emission near the nucleus ($R<2.5''$), with velocities exceeding 100 km/s. This molecular gas is outflowing, not participating in disc motions, and it has been removed from the moment-0 map used to estimate the molecular gas distribution in the disc.}
\label{fig:pvdiagram}
\end{figure}

\subsection{Harmonic decomposition of the gravitational potential} 
\label{Sec:harmonicpot}

To assess the radial contribution of the different Fourier modes, we performed a harmonic decomposition of the potential,

\begin{equation}
\Phi (R, \theta) = \Phi_0 (R) + \sum_m \Phi_m (R) \cos(m\theta - \phi_m(R))
,\end{equation}

\noindent
where $\Phi_m (R)$ and $\phi_m (R)$ are the amplitude and phase of the $m$-mode.

We followed \citet{1981A&A....96..164C} and calculated the strength of the $m^\mathrm{th}$ Fourier component, $Q_m (R)$, as

\begin{equation}
Q_m (R) = \frac{m \Phi_m (R)}{R |F_0 (R)|},
\end{equation}

\noindent
which is the amplitude of the $m^\mathrm{th}$ harmonic component normalised by the mean axisymmetric radial force, $F_0 (R)$. Finally, we also computed the maximum tangential force in terms of the mean radial force,

\begin{equation}
Q_\mathrm{T} (R) = \frac{F\mathrm{_T^{max}}(R)}{F_0 (R)}=
\frac{\frac{1}{R} \left(\frac{\partial \Phi(R,\theta)}{\partial \theta} \right)_\mathrm{max}} {\frac{\mathrm{d} \Phi_0(R)}{\mathrm{d} \theta}}.
\end{equation}

\begin{figure}[t]
\begin{center}
\includegraphics[trim=16 280 -10 0,clip,width=0.48\textwidth]
{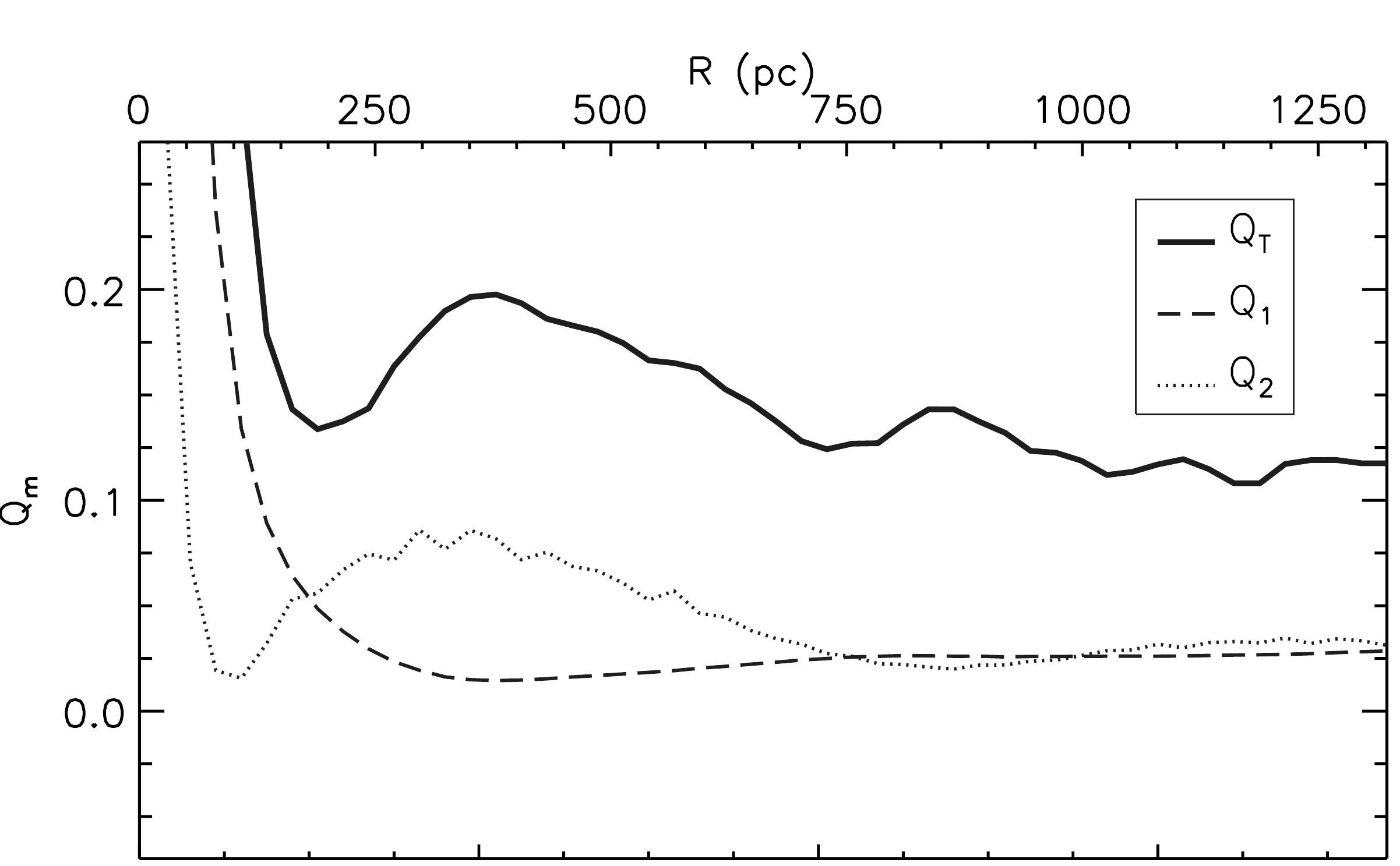}
\includegraphics[trim=16 250 -10 30,clip,width=0.48\textwidth]
{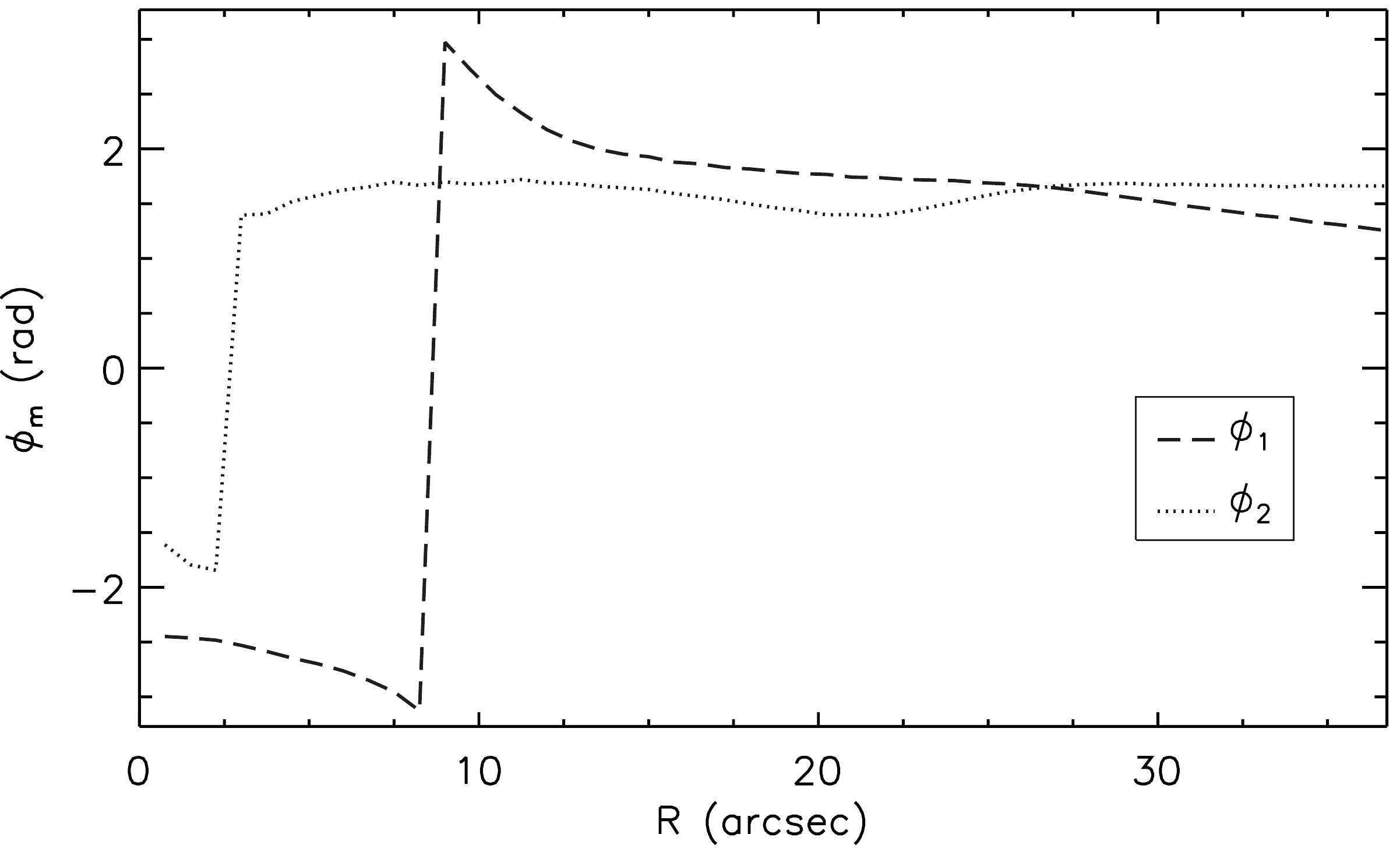}
\end{center}
\caption{\textit{Top:} relative strength of the first two Fourier modes in the 
gravitational potential ($Q_{m=1,2}$) as a function of radius, compared to the total strength of the non-axisymmetric perturbation $Q_\mathrm{T}$. \textit{Bottom:} phases of the first two Fourier modes in the gravitational potential $\phi_{m=1,2}$, measured counter-clockwise from the major axis of the galaxy on the deprojected image (represented here in radians from $-\pi$ to $\pi$).}
\label{fig:harmonicpot}
\end{figure}

The top panel of Fig.\,\ref{fig:harmonicpot} shows the relative strength of the first two modes ($m=1,2$) as a function of radius, compared to the total strength of the non-axisymmetric perturbation $Q_\mathrm{T}$; the bottom panel shows the corresponding phases, $\phi_{m=1,2}$. These results are discussed in Sect.\,\ref{Sec:stellarpot}.

\subsection{Torques from the potential} 
\label{Sec:torques}

We obtained the forces per unit mass as the gradient of the potential in Cartesian coordinates,

\begin{equation}
F_{x,y}(x,y) = -\nabla_{x,y} \Phi(x,y),
\end{equation}

\noindent
and, subsequently, we used these forces (the components of the local force along each of the Cartesian axes) to obtain the torques,

\begin{equation}
\tau(x,y) = x F_y - y F_x.
\end{equation}

It is important to emphasise that these are torques per unit mass (i.e. torques per unit gas mass, since we are interested
in stellar gravitational torques on the gaseous component), and we measured them in units of $\mathrm{km}^{-2} \, \mathrm{s}^2$. Therefore, multiplying the $\tau(x,y)$ in a given pixel by the total gas mass contained in that pixel (e.g. traced by CO) will provide the change rate of angular momentum with respect to time experienced by the gas ($\tau=dL/dt$).

\begin{figure*}[t]
\begin{center}
\includegraphics[trim=0 210 0 0,clip,width=1.0\textwidth]{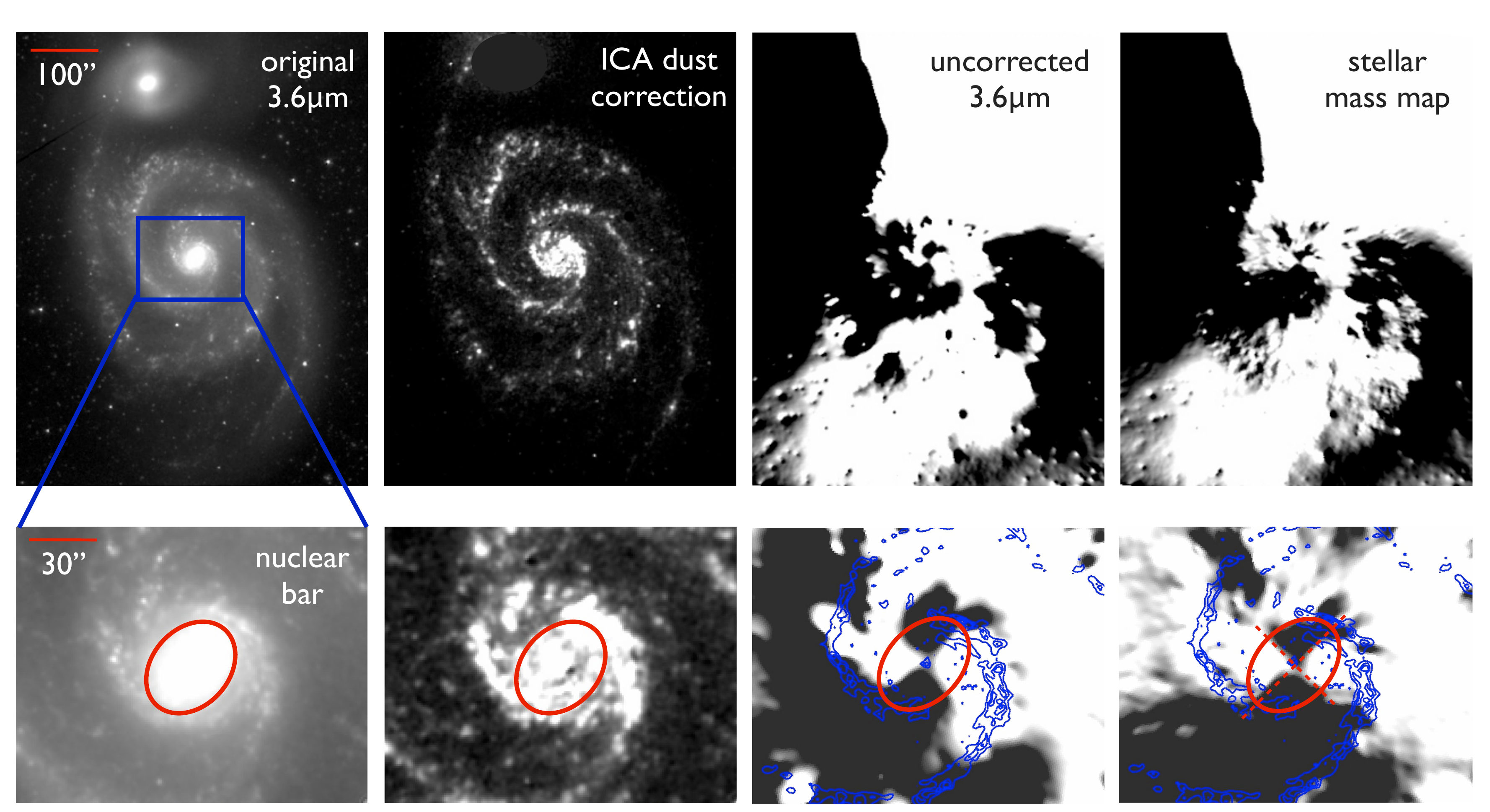}
\end{center}
\caption{\textit{Top panel:} original 3.6\,$\mu m$ image of M51; dust emission identified with ICA (subtracted from the original image to produce the stellar mass map); gravitational torques, $\tau(x,y)$, based on the uncorrected 3.6\,$\mu m$ image of M51 (assuming a constant M/L); same gravitational torques, $\tau(x,y)$, based on our ICA dust-corrected stellar mass map of M51. 
\textit{Bottom panel:} blow-up of the maps above, showing the area close to the nucleus. The red ellipse displays the shape of the nuclear bar, determined from the the ellipticity and PA profiles measured by \citet{2015ApJS..219....3M} at the radius where the bar ends ($R=20''$: $\varepsilon=0.262$, $PA=-42.6^\circ$); this agrees excellently well with an isocontour of $F=10$\,MJy/sr in the original 3.6\,$\mu m$ image. The red dashed lines are the approximate axes of symmetry of the bar, which coincide with a change of sign in the torques, following the expected butterfly pattern. The blue contours indicate CO emission from PAWS, from 50 to 250\,K\,km\,s$^{-1}$ in steps of 50\,K\,km\,s$^{-1}$. All images are shown in the plane of the sky; North is up (even if, naturally, torques are calculated in the plane of the galaxy). The scale bar shown in the leftmost images applies to the whole row.
[The 3.6\,$\mu m$ image is shown in square-root scale to emphasise low-level structure in the range $(0,10)$\,MJy/sr; the dust map is displayed in linear scale, $(0,3)$\,MJy/sr; the torque maps are also shown on a linear scale, $(-1000,1000)$\,km$^2$\,s$^{-2}$, with negative torques in black and positive torques in white.]}
\label{fig:depro-torques}
\end{figure*}

Figure~\ref{fig:depro-torques} 
shows the gravitational potential and the torques that stem from our ICA-corrected stellar mass map, deprojected according to the PAWS measurements of PA and ellipticity.

\subsection{Gas flows} 
\label{Sec:gas}

Torques applied on gas are, by definition, a measure of the change of angular momentum per unit time ($\tau=dL/dt$). Naturally, angular momentum loss of the gas is associated with inflow, whereas increasing its angular momentum corresponds to radial outflow. Multiplying the torque map by the present-day gas distribution (e.g. from PAWS) provides the instantaneous view of angular momentum loss and gain across the galaxy. However, this is not necessarily representative, overall, of the net effect experienced by a given gas cloud. We followed \citet{2005A&A...441.1011G} and computed an azimuthal average of the torques, weighted with the local gas column density, $N(x,y)$,

\begin{equation}
\tau(R) = \frac{\int_\theta[N(x,y) \cdot (xF_y-yF_x)]}{\int_\theta N(x,y)}.
\end{equation}

To weight the torques, we used the PAWS CO(1-0) map of M51 corrected for outflow motions (Sect.\,\ref{Sec:PAWS}) and convolved to the spatial resolution of the stellar mass map (PSF$_{3.6 \mu m }=1.7''$). At $R=4$\,kpc, HI starts to dominate \corr{over} CO \citep{2007A&A...461..143S},
therefore we used HI traced by THINGS to extend the profiles to larger radii. We deprojected the gas maps using the same parameters as for the stellar mass map, following Sect.\,\ref{Sec:depro}.

To calculate the fuelling efficiency, we normalised the azimuthally averaged torques $\tau (R)$ at each radius by the angular momentum and rotation period. This provides an estimate of the average fraction of gas specific angular momentum transferred by the stellar potential in one rotation ($T_\mathrm{rot}$),

\begin{equation}  \label{eq:AL_L}
\frac{\Delta L}{L} = \left. \frac{dL}{dt} \right|_\theta \cdot \left. \frac{1}{L} \right|_\theta \cdot T_\mathrm{rot} = \frac{\tau (R)}{L_\theta} \cdot T_\mathrm{rot},
\end{equation}

\noindent
since $\tau (R)$ is, by definition, the azimuthally averaged time-derivative of the specific angular momentum of the gas. The azimuthal average of the angular momentum was assumed to be $L_\theta = R \cdot v_\mathrm{rot}$. The inverse of Eq.~\ref{eq:AL_L}, $L / \Delta L$, represents the number of rotation periods needed to transfer all the angular momentum of the gas; a large $L / \Delta L$ would imply that the gas distribution and potential do not vary too much on the timescale of a rotation, which justifies the approximation introduced above. Finally, based on this, we express the gas inflow/ouflow rate per unit length as 

\begin{equation}  \label{eq:d2MdRdt}
\frac{d^2M(R)}{dRdt} = \left. \frac{dL}{dt} \right|_\theta \cdot \left. \frac{1}{L} \right|_\theta \cdot 2 \pi R \cdot N(x,y)|_\theta,
\end{equation}

\noindent
which provides the result in units of $M_\odot \mathrm{yr}^{-1} \mathrm{pc}^{-1}$. This can be integrated radially, using radial bins of width $\Delta r$, to obtain a net inflow/outflow rate up to a given radius $R$,

\begin{equation}  \label{eq:dMdt}
\frac{dM(R)}{dt} = \sum_{r=0}^{R} \left [\frac{d^2M(r)}{drdt} \cdot \Delta r \right].
\end{equation}

\begin{figure*}[t]
\begin{center}
\includegraphics[trim=30 400 0 0,clip,width=0.92\textwidth]{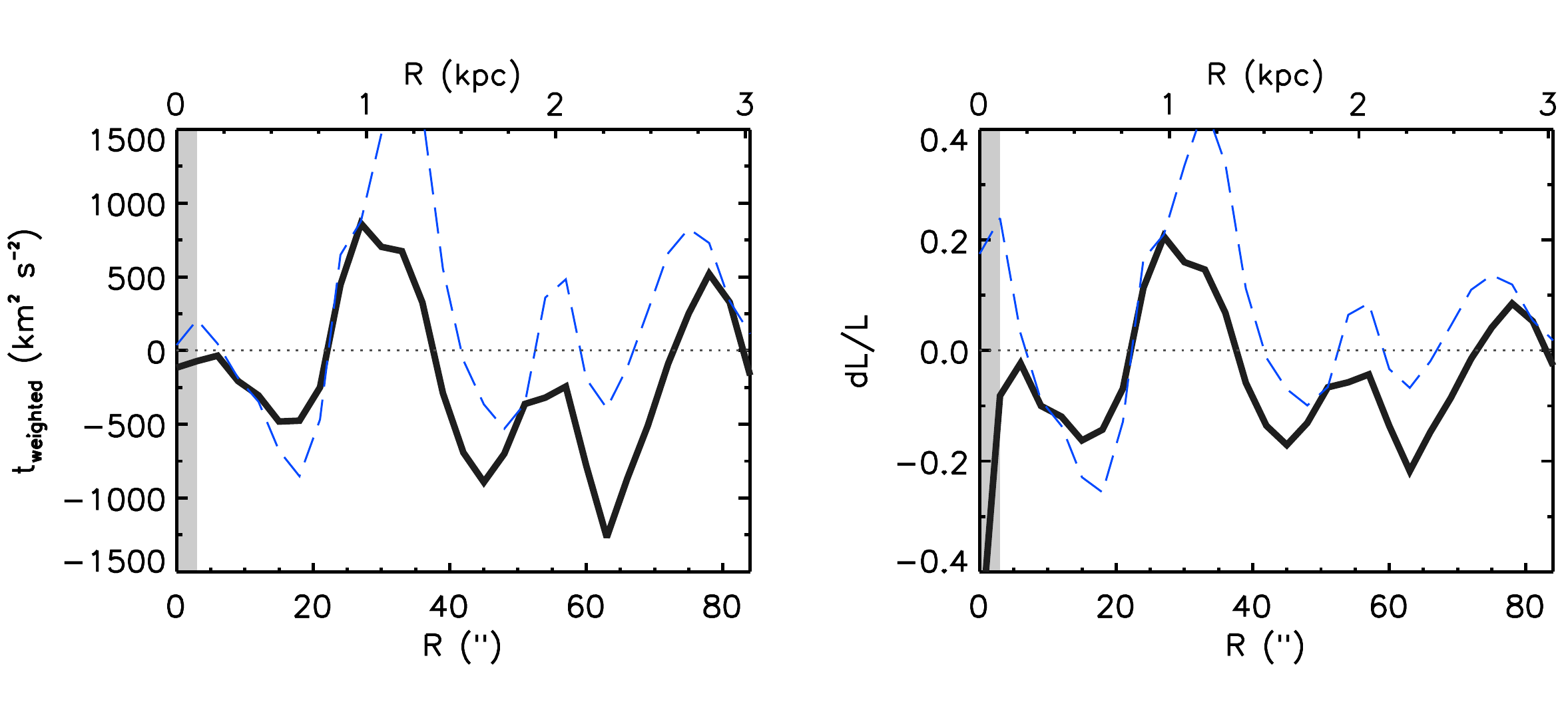}
\includegraphics[trim=30 420 0 0,clip,width=0.92\textwidth]{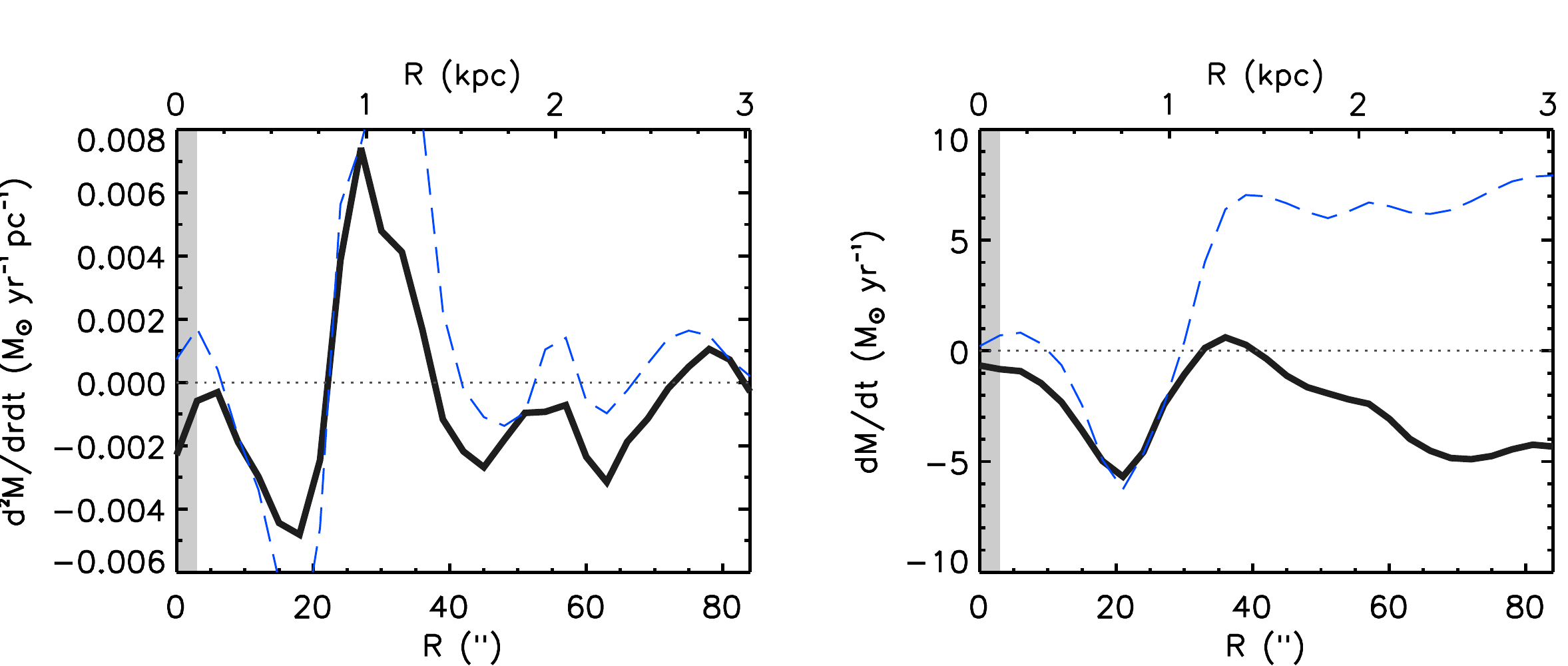}
\end{center}
\caption{Results of the gravitational torques (weighted with the molecular gas distribution), angular momentum transfer, and resulting gas flow, both in radial bins and integrated out to radius $R$ for our ICA dust-corrected stellar mass map (solid black line) and the original 3.6\,$\mu m$ image (dashed blue line). Positive values imply radial gas outflow, negative values denote an inflow. The innermost bin has been shaded to emphasise that it is subject to high uncertainties.}
\label{fig:profiles}
\end{figure*}

If measured close enough to the nucleus, $dM/dt$ in Eq.\,\ref{eq:dMdt} provides an estimate of the instantaneous AGN feeding rate. Figure
\,\ref{fig:profiles} presents these results.

\subsection{Subtracting the outflow component} 
\label{Sec:outflow}

Our method to estimate gas flows based on the gravitational torques assumes planar motions for the molecular gas. The regular kinematics of molecular gas in M51 and the relatively constant CO line width \citep{2013ApJ...779...43P} are indicative that this assumption is reasonable. However, near the nucleus, significant molecular emission is clearly deviating from disc motions: this has been argued to be a molecular gas outflow \citep{1998ApJ...493L..63S,2004ApJ...616L..55M,2007A&A...468L..49M,2015ApJ...799...26M}, and it becomes clear in the position-velocity diagram shown in Fig.\,\ref{fig:pvdiagram}.
To correct for it, we subtracted the CO(1-0) emission that is not associated with disc motion by constructing a new moment-0 map \citep[using 3-$\sigma$ clipping,][]{2013ApJ...779...43P} in which we neglected the flux from pixels in the central $R=2.5''$ in channels of velocities that correspond to the outflow (40-180 km/s).
 We verified that the final torque profiles do not change significantly if this cut is modified by $\pm$10~km/s; this implies that our
model is reliable.

\section{Results}
\label{Sec:results}

\subsection{Stellar potential}
\label{Sec:stellarpot}

Figure\,\ref{fig:harmonicpot} shows the $Q_m$ values for the first two Fourier modes ($m=1,2$) in the stellar gravitational potential of M51. These represent the strength of each mode in the potential as a function of radius (compared to the total non-axisymmetric contribution, $Q_\mathrm{T}$). There is a clear dominance of the $m=2$ mode all the way out to $R = 20''$, which can be attributed to the influence of the nuclear bar; indeed, the bar has a length of $R_\mathrm{bar} \approx 20''$ \citep[][]{2010MNRAS.402.2462C}. The area of influence of the
bar is therefore similar to its size as measured on NIR images, as expected.

The bar is considerably strong, reaching a maximum $Q_2 \sim 0.1$ at a radius of $10''$ $\sim$ 400\,pc.
In addition, the phase of the $m = 2$ bar in Fig.\,\ref{fig:harmonicpot}, which we find is relatively constant around $\phi_2 \sim \pi/2$ radians (oscillating between 1.5 and 1.7, with an average of 1.63 radians), corresponds to a disc-plane bar orientation\footnote{In our Fourier expansion we assumed positive amplitudes, $\Phi_m (R) \cos[m\theta - \phi_m(R)]$ with $\Phi_m (R) >0$, which implies that the major axis of the bar, where the potential is lowest, is given by $2\theta - \phi_2 = \pi$ (or an integer multiple of $\pi$). Therefore, $\theta = \pi / 2+\phi_2 / 2 \sim 3\pi / 4 = 135^\circ$ on the plane of the galaxy.} of 135$^\circ$ measured counterclockwise from the major axis of M51 or, equivalently, a sky-plane orientation of PA$ = -47^\circ$ (given the projection of the disc). This agrees well with the PA value of $-43^\circ$ given by the profiles of \citet{2015ApJS..219....3M} at a radius of $R=20''$ and the bar angle of $-41^\circ$ measured by \citet{2007ApJ...657..790M} from NIR imaging.

\subsection{Torque map of M51}
\label{Sec:torquemap}

Figure\,\ref{fig:depro-torques} shows the torques implied by the stellar mass map. In all these maps, black corresponds to negative torques and white implies positive torques; with our sign convention (right-hand rule), negative torques make the gas lose angular momentum (inflow), while positive torques transfer additional angular momentum to the gas (outflow). The blow-up shown in the bottom panels demonstrates the good agreement between the morphology of the nuclear bar (delineated in red) and the butterfly pattern in the torque map: torques change signs according to four quadrants that are delimited by the axes of symmetry of the bar, as expected. The nuclear bar extends out to a radius of $R=20'' \sim 800$\,pc \citep{2010MNRAS.402.2462C}. As we show
in Sect.\,\ref{Sec:inflowrates}, there is evidence for gas inflow inside $R=22'' \sim 880$\,pc, which, judging from the butterfly pattern visible in the torque map, can be clearly interpreted as the effect of the bar.

The right panels of Fig.\,\ref{fig:depro-torques} stress the difference with the torque map stemming from the uncorrected 3.6\,$\mu m$ band (i.e. stellar mass distribution obtained multiplying the 3.6\,$\mu m$ image with a single M/L); the change is even more evident in the radial profiles in Fig.\,\ref{fig:profiles}. The presence of dust emission leads to a significantly different distribution of torques, which results in no consistent gas inflow in the central area (in spite of the morphological evidence for a nuclear bar). One of the side-effects of correcting the 3.6\,$\mu m$ image for dust emission using ICA is that small oversubtractions appear in areas of very red original [3.6]-[4.5] colours, which manifest themselves as small holes or dipoles in the torque map. However, in Appendix\,\ref{Sec:oversubtr} we show that these small oversubtractions due to the ICA correction only have a
weak effect when they are azimuthally averaged, while the difference with the uncorrected 3.6\,$\mu m$ image is large.

In addition to the butterfly pattern associated with the nuclear bar, the torque maps in Fig.\,\ref{fig:depro-torques} display a second butterfly pattern at larger radii, which is tilted with respect to the nuclear \textcolor[rgb]{1,0.501961,0}{} {butterfly
pattern }(especially obvious at $R \gtrsim 80'' \sim 3$\,kpc).
Comparison to a smooth mass model of M51 demonstrates that this is probably due to an \textup{\textit{\textup{oval distortion in the disc}}.} For a perfectly axisymmetric (circular) disc, no butterfly pattern is expected; but in the case of M51, the disc is photometrically elongated along an axis of $PA=39^\circ$ \citep{2015ApJS..219....4S}, while kinematics imply that the orientation of the (molecular) disc is such that $PA=172^\circ$ \citep{2014ApJ...784....4C}. Provided that the disc is not significantly warped inside $r \lesssim 6$\,kpc \citep{2014ApJ...784....4C}, the most natural interpretation for such a configuration would be that the disc is not circular, but rather oval, which might be expected as a consequence of the interaction 
\citep[a global $m=2$ mode induced by the encounter with NGC\,5195; see also][]{2013ApJ...779...45M}. Simulations from \citet{2010MNRAS.403..625D} also support the idea that the disc is not strongly warped inside 5\,kpc and that there is an oval at $r \sim 2$\,kpc, with aspect ratio 2:1 \citep[see also][on the possibility of a warp in M51]{2007ApJ...665.1138S}.

\subsection{Inflow rate in M51}
\label{Sec:inflowrates}

Our best estimate of the gas flows is presented in Fig.\,\ref{fig:profiles}. The top-left panel shows that the (gas-weighted) azimuthally averaged torques are consistently negative from the centre all the way out to $R=22''$, which implies secular molecular gas inflow in the central $\sim 1$\,kpc of the galaxy \referee{(down to our resolution limit of $1.7''=63$\,pc)}. After this, in the range $R=(22-37)''$, torques become positive, which is indicative of radial molecular gas outflow along the plane of the galaxy; then, from $R=37''$ to $R=73''$, torques are again continuously negative, implying inflow. This means that there is evidence for secular transport of the gas from $R=2.7$\,kpc 
\referee{down to our resolution limit}
 with the exception of the range $R=(22-37)''$, which seems to be a dynamical barrier for gas transport; precisely in this region
the molecular ring is located \citep[$R \sim 30''$][]{2013ApJ...779...45M}, which can be explained as the accumulation of molecular gas as a consequence of these gas flows \citep[e.g.][]{1996FCPh...17...95B,2013A&A...556A..98V}. 
\referee{This seems to be the same structure as the inner pseudo-ring identified by \citet{2014A&A...562A.121C} around $R \sim 20''$ \citep[which is star-forming,][]{2013A&A...555L...4C}. It becomes especially obvious in the dust component separated by the ICA correction of the 3.6\,$\mu m$ image (second panels of Fig.\,\ref{fig:depro-torques}, at the end of the ellipse that delineates the bar).}
 As a transition region between the nuclear bar and the radius where spiral arms end, the molecular ring is not as strong and dominant as rings in other galaxies; in this sense, it can be associated with the zero-crossing occurring around $R \sim 40''$ in the torque profile, which coincides with the radius at which the spiral arms terminate.

The top-right panel represents the fraction of the angular momentum of the gas at a certain radius that is lost in one rotation. The fact that $dL/L$ takes a maximum value of the order of 0.2 (20\%) justifies the assumption (implicit to azimuthal averaging) that the gas distribution does not change by much in one rotation.  
\referee{Out to} $R=22''$, ignoring the innermost bin, the average dL/L is $\sim 10\%$, which means that the gas will lose all of its angular momentum on a timescale of $\sim 10$ rotations, that
is, in about 200\,Myr ($T_\mathrm{rot} (R=12'')=22$\,Myr). This timescale is also relatively short compared to the dynamical timescale of the interaction of M51 with NGC\,5195 ($\sim$1\,Gyr), which allows us to assume that the gravitational potential of M51 will not have changed by much in that period.

The bottom \referee{left} panel of Fig.\,\ref{fig:profiles} shows the instantaneous flow rates implied by the gravitational torques across the disc of the galaxy, the bottom right panel shows the `integrated' inflow rates (i.e. summing the contributions of all bins out to radius $R$). The `integrated' inflow rate out to $R = 22''$ is $\sim 5\,M_\odot /yr$, which means that on a timescale of one year, a total of about $5\,M_\odot$ of gas participates in \textit{net} inflow motions in that particular radial extent of the galaxy. Closer to the AGN (considering only the first or first two bins, 3-6”), this rate decreases to $\sim 1\,M_\odot /yr$, with the caveat that the innermost bin is compromised by the outflow and by increasingly large systematic uncertainties (see Sect.\,\ref{Sec:uncert}).
\refereetwo{In Appendix\,\ref{sec:F190N} we confirm these results using the gravitational potential computed from an HST 1.9\,$\mu m$ image (instead of the stellar mass map). The inflow rates and resonances are basically coincident with those presented here, and significant divergencies only appear in the inner $R \lesssim 5''$, where uncertainties become increasingly large and data are becoming inconclusive as a result of the additional impact of the central AGN.}

\subsection{Dynamical resonances in M51}
\label{Sec:resonances}

We used our new map of gravitational torques to reassess the locations of the main dynamical resonances in the disc of M51.   
\citet{2013ApJ...779...45M} have used the technique of constructing radial profiles of the azimuthally averaged torques to assess the positions of resonances in M51 with PAWS. Here we also used a 3.6\,$\mu m$-based stellar mass map (distinct from the earlier version of \citealt{2013ApJ...779...45M}, as discussed in Sect.\,\ref{Sec:Discussion}), but we extended the analysis out to larger radii using the more extensive area probed by the HERACLES and THINGS maps of the molecular and atomic gas distributions \citep{2008AJ....136.2563W,2009AJ....137.4670L}. This avoids potential biases due to the non-circular shape of the PAWS field of view, which leads to incomplete azimuthal coverage at radii $R>85''$.

\begin{table}[t!]
\begin{center}
\caption[h!]{Dynamical resonances of M51 from our torque analysis.}
\begin{tabular}{lcc}
\hline\hline 
Resonance               &       Radius  ('') &  Radius  (pc)    \\
\hline 
Bar corotation &         $22 \pm 2.5$   &       $830 \pm 90$    \\
Bar outer Lindblad resonance    &       $50 \pm 5$      &       $1800 \pm 180$        \\
Spiral corotation       &       $102 \pm 10$    &       $3800 \pm 370$        \\
Spiral ultra-harmonic resonance &       $55 \pm 5$      &       $2000 \pm 180$                \\
Spiral outer Lindblad resonance &       $165 \pm 15$ &  $6100 \pm 550$        \\
 \hline
\end{tabular}
\label{table:reson}
\end{center}
\end{table}

In the framework of the density wave theory, the response of gas to the gravitational potential is expected to change at 
corotation (CR). The torque profile obtained at highest angular resolution using the PAWS map exhibits an abrupt change of sign (from negative to positive torques) at $R=22''$, which can be associated with the CR of the nuclear bar \citep[$R_\mathrm{CR,bar}$; see also][]{2013ApJ...779...45M,2014ApJ...784....4C}. \citet{2010MNRAS.402.2462C} obtained a bar radius of $R_\mathrm{bar}=20''$ (800\,pc) based on ellipse fitting, which implies a ratio of $R_\mathrm{CR,bar}/R_\mathrm{bar} \approx 1.1$. This agrees with expectations for fast bars \citep[e.g.][]{1992MNRAS.259..328A} and is also well compatible with the values measured by \citet{2008MNRAS.388.1803R} for galaxies of similar morphological type (SABbc).
\referee{As we commented above, the inner star-forming ring identified by \citet{2014A&A...562A.121C} around $R \sim 20''$ overlaps the molecular ring and is located next to the corotation of the bar, suggesting that it is indeed a resonant ring.}

\begin{figure}[H]
\center
\includegraphics[trim=0 70 0 0,clip,width=0.48\textwidth]{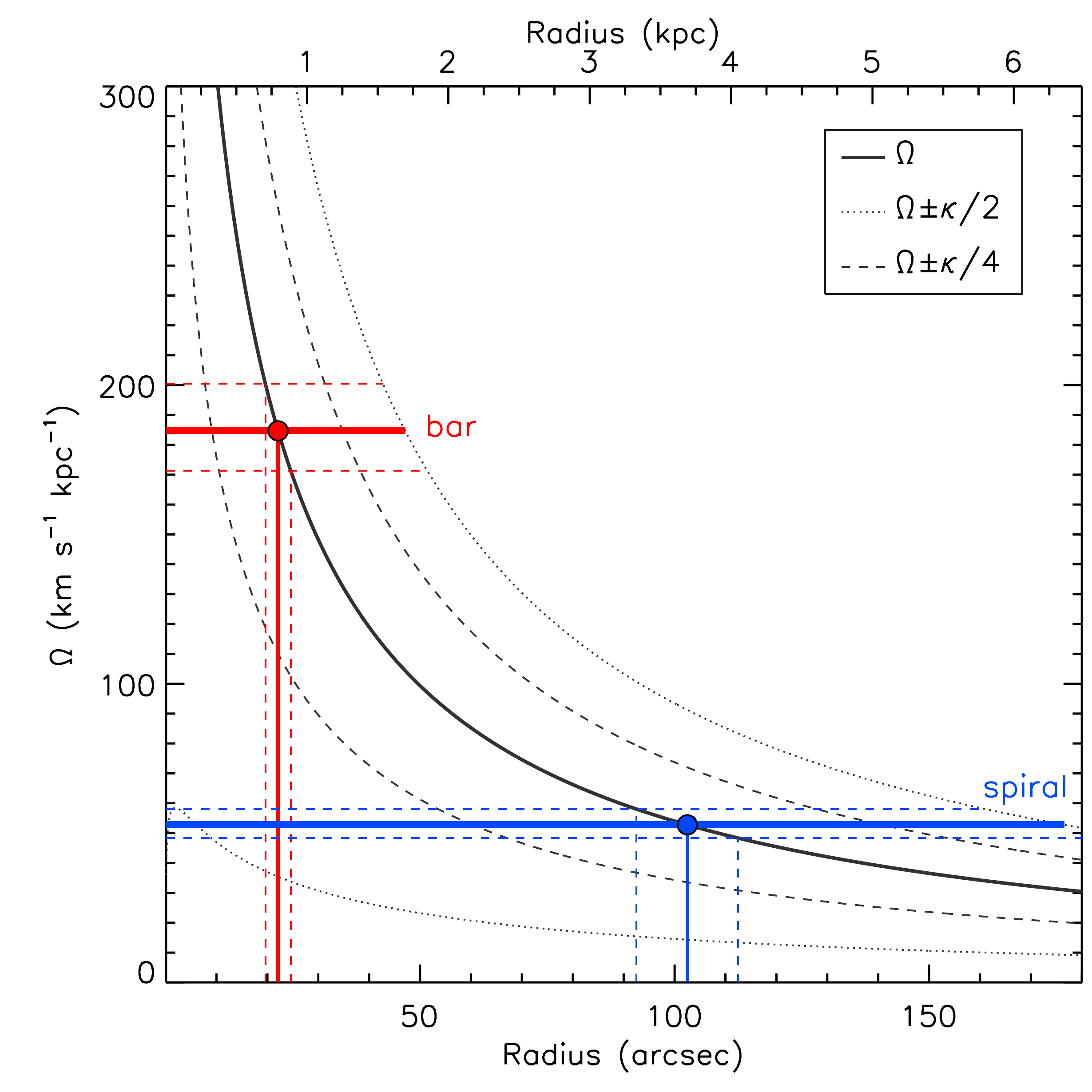}
\includegraphics[trim=0 315 0 30,clip,width=0.48\textwidth]{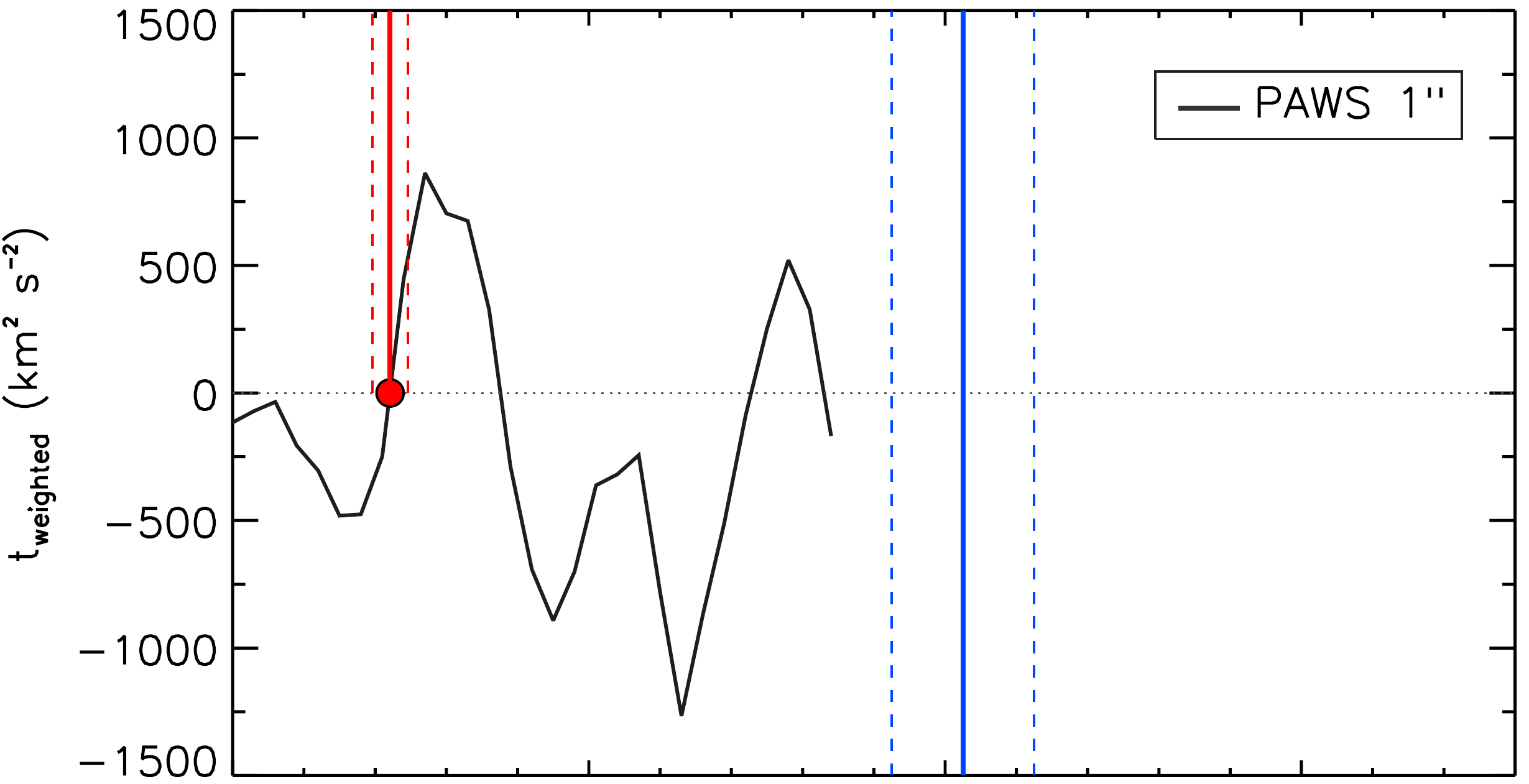} 
\includegraphics[trim=0 250 0 32,clip,width=0.48\textwidth]{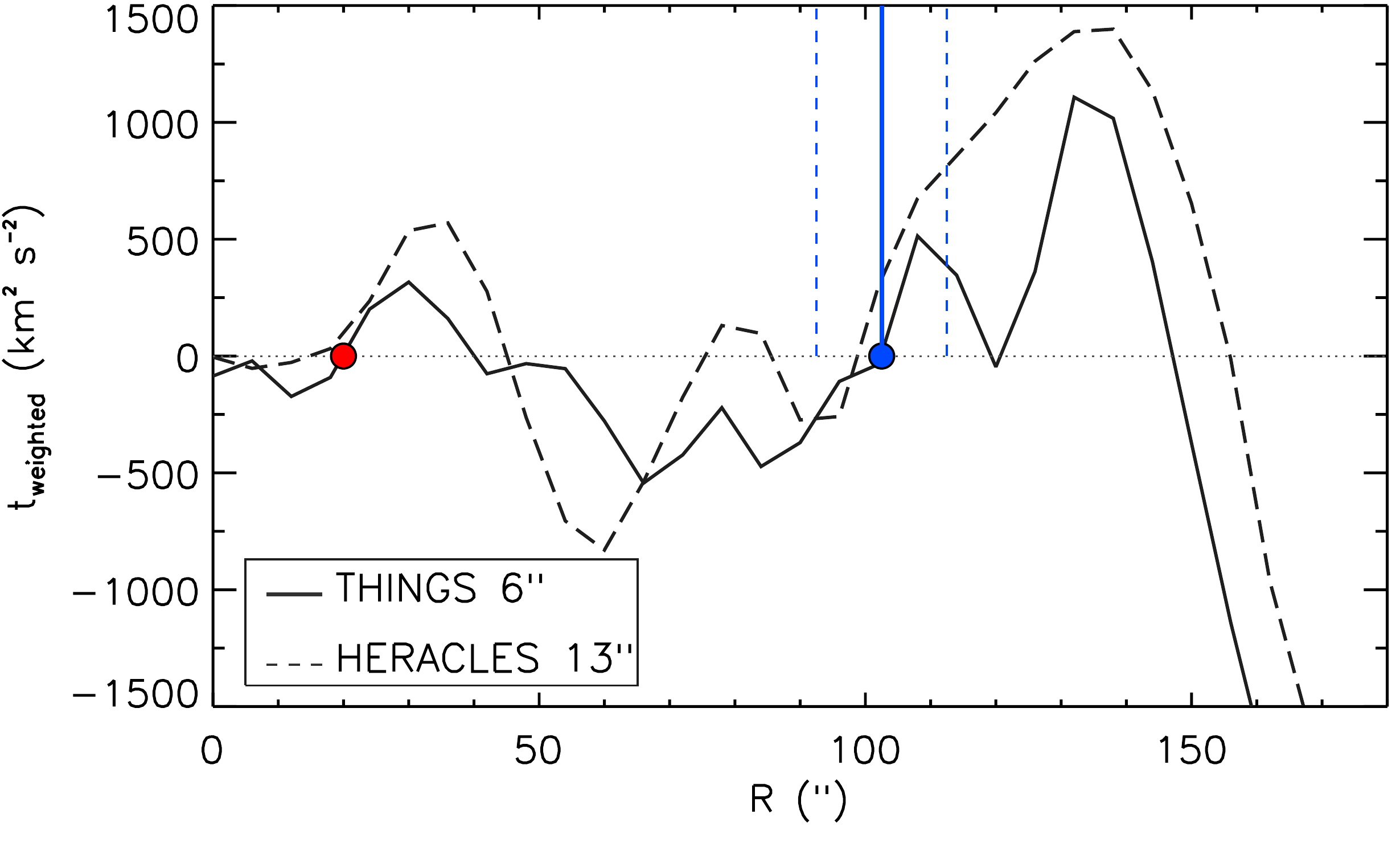} 
\caption{\textit{Top panel:} Angular frequency plot based on the rotation curve of M51 calculated in \citet{2013ApJ...779...45M}: $\Omega$ (solid black), $\Omega \pm \kappa / 2$ (dotted black), $\Omega \pm \kappa / 4$ (dashed black). The red circle indicates the crossing of the estimated corotation radius of the bar (CR$_\mathrm{bar}=(22 \pm 2.5)''$) with the angular frequency curve $\Omega$, which implies a bar pattern speed of $\Omega_P^\mathrm{bar} =(185 \pm 15)$\,km\,s$^{-1}$\,kpc$^{-1}$ (horizontal red line). The blue circle shows the analogous crossing of the suggested corotation radius of the spiral (CR$_\mathrm{sp}=(102 \pm 10)''$), implying a pattern speed for the spiral of $\Omega_P^\mathrm{spiral} =(53 \pm 5)$\,km\,s$^{-1}$\,kpc$^{-1}$ (horizontal blue line).
\newline \textit{Middle panel:} Torque profile weighted by the PAWS gas distribution (same as Fig.\,\ref{fig:profiles}, reproduced here for comparison). As a result of the smaller field of view of PAWS, the profile stops at $R=85''$.
\newline \textit{Bottom panel:} Torque profile weighted by the gas emission traced by THINGS (solid line) and HERACLES
(dashed line).}
\label{fig:angfreq}
\end{figure}

Inner torque variations locating the bar corotation are not as obvious in the lower resolution profiles obtained with THINGS\footnote{Additionally, HI is almost completely depleted in the inner regions.}
 or HERACLES (the latter of which covers the bar with only two resolution elements). However, we expect these maps to offer a more reliable measure of gravitational torques at larger radii, where the gravitational torques can be estimated with full (0-2$\pi$) azimuthal information. 
A second zero-crossing from negative to positive torques is visible in the THINGS profile near $100''$ (Fig.\,\ref{fig:angfreq}, bottom panel), which we suggest is indicative of corotation for the spiral pattern; this interpretation is reinforced by a similar crossing in the HERACLES profile (with an offset of only $5''$, i.e. fully compatible given the HERACLES resolution of $13''$). The final change of sign at $R \sim 150-160''$ probably marks the outer Lindblad resonance (OLR) of the spiral.  

These findings qualitatively agree with the results of the harmonic decomposition of the PAWS line-of-sight velocity field performed at $1''$ resolution by \citet{2014ApJ...784....4C}. These authors observed a change in dominance from the $s1$ to the $s3$ harmonic coefficients around $R \sim 100''$, which we associate with the location of corotation.
As first shown by \citet{1993ApJ...414..487C}, a change from $m=1$ to $m=3$ mode is expected at corotation, which will reflect on a change in dominance of the $s1$ and $s3$ terms of the Fourier harmonic decomposition of the velocity field 
\citep[even if the behaviour of those harmonic terms is more complex in spirals than in bars; e.g.][]{2004ApJ...605..183W}.
We strongly recommend combining the information contained in torque profiles with other (kinematic and morphological) evidence to accurately identify resonances. In particular, while corotation
is expected to result in a zero-crossing in the torques, 
 other resonances might also have an imprint in the form of a change of sign. 
Our estimates for the bar and spiral corotation radii in M51 are also consistent with the values independently estimated by \citet{2012arXiv1203.5334Z} ($R_\mathrm{CR1}=25''$ and $R_\mathrm{CR2}=110''$).

A very useful tool to interpret the results on dynamical resonances and analyse possible couplings is a so-called angular frequency plot, in which the angular frequency corresponding to circular rotation is plotted ($\Omega = V_\mathrm{rot} / R$) along with the epicyclic curves (e.g. $\Omega \pm \kappa / 2$, $\Omega \pm \kappa / 4$). In Fig.\,\ref{fig:angfreq}, we show those curves as derived from the three-parameter rotation curve of M51 calculated in \citet{2013ApJ...779...45M}.
For a direct comparison, the middle and bottom panels of Fig.\,\ref{fig:angfreq} show the torque profiles (weighted by PAWS and THINGS/HERACLES, respectively) and some coloured circles indicating the corotation positions that we have inferred (red for the bar, blue for the spiral). The intersections of the zero-cossings with the angular frequency curve $\Omega$ in the top plot indicate the corresponding pattern speeds (since $\Omega (R_\mathrm{CR}) = \Omega_\mathrm{P}$). The dashed coloured lines show the range of possible values due to 1.5 times the resolution of each survey (2.5$''$ for PAWS; 10$''$ for THINGS), in an attempt to give an idea of how those observational uncertainties translate into a range of possible pattern speeds and overlapping resonances.

 The pattern speed of the bar (for a CR$_\mathrm{bar}=(22 \pm 2.5)''$) would be $\Omega_P^\mathrm{bar} =(185 \pm 15)$\,km\,s$^{-1}$\,kpc$^{-1}$ (horizontal red line). Analogously, for our suggested corotation radius of the spiral (CR$_\mathrm{sp}=(102 \pm 10)''$), the spiral pattern would rotate with an angular frequency of $\Omega_P^\mathrm{spiral} =(53 \pm 5)$\,km\,s$^{-1}$\,kpc$^{-1}$ (horizontal blue line). Placing the main corotation of the spiral around $R \sim 100''$, as suggested by our torque analysis, has interesting implications in terms of coupled resonance patterns. On the one hand, if a single (or dominant) pattern speed applies to the whole spiral, the crossing of $\Omega_P^\mathrm{spiral} =(53 \pm 5)$\,km\,s$^{-1}$\,kpc$^{-1}$ with the $\Omega + \kappa / 2$ curve would imply an OLR$_\mathrm{sp} \sim (150 - 180)''$, in agreement with torques (last zero-crossing around $(150 - 160)''$). The inner ultraharmonic resonance of the spiral (UHR, or 4:1) is given by the crossing of that same pattern speed with the $\Omega - \kappa / 4$ curve, which results in UHR$_\mathrm{sp} \sim (50 - 60)''$; this is compatible with the OLR of the bar ($\Omega + \kappa / 2$ curve crossing with $\Omega_P^\mathrm{bar} \sim 185$\,km\,s$^{-1}$\,kpc$^{-1}$), which points to a possible dynamical coupling of the two structures \citep[on resonance coupling, see e.g.][]{1987ApJ...318L..43T,2014ApJS..210....2F}. These measurements are summarised in Table\,\ref{table:reson}.
 
 Determining the location of the ILR is often more challenging in the discs of spiral galaxies, given the characteristic rapid changes in the shape of the innermost parts of rotation curves. In M51, the shape of the $\Omega - \kappa / 2$ curve at radii $r \lesssim 30''$ is very sensitive to the exact parameterisation of the rotation curve we adopt.  \referee{Given this uncertainty in the inner shape of the $\Omega - \kappa / 2$ curve, we note that the resonant overlap of the bar CR and spiral ILR might be possible (although this is difficult to claim with great certainty); this would further support the resonant nature of the inner pseudo-ring around $R \sim 20''$.}

Based on all this, 
 we propose a scenario in which M51 has two main corotations: one associated with the nuclear bar, approximately at CR$_\mathrm{bar} \sim 20''$, and one for the spiral around CR$_\mathrm{sp} \sim 100''$. There are only two clear zero-crossings (from negative to positive) in the extended torque profiles using THINGS and HERACLES. These are consistent with the $s1/s3$ radial profiles from the harmonic decomposition of the PAWS and THINGS velocity fields.

\subsection{Use of a dust correction map to estimate flows and resonances}
\label{Sec:s2map}

\begin{figure*}[t]
\begin{center}
\includegraphics[trim=0 420 0 0,clip,width=1.0\textwidth]{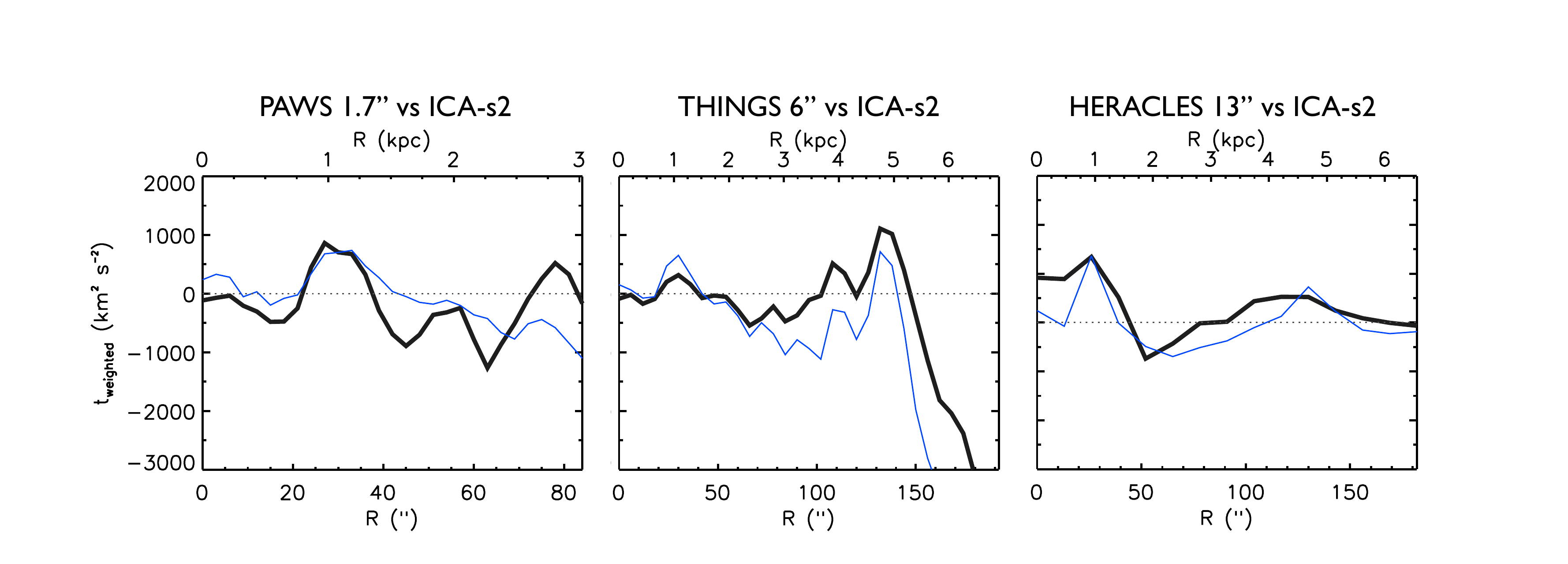}
\end{center}
\caption{Torque profiles using PAWS (1.7$''$ resolution), THINGS (6$''$), and HERCALES (13$''$) compared to the equivalent profiles using the ICA $s2$ component (dust map) at matched resolution. The thick black line is the fiducial profile (using the actual gas distribution), the thin blue line corresponds to $s2$. The agreement is not one-to-one, but $s2$ can be used as a first-order approximation in the absence of gas information at high enough resolution.}
\label{fig:s2}
\end{figure*}

In Sect.\,\ref{Sec:stellarmass} we outlined the interest of correcting the 3.6\,$\mu m$ image for dust emission using an ICA technique. Here, we briefly examine the possibility of using the dust correction map ($s2$) produced by the ICA separation as a tracer of the ISM. The $s2$ map is uniformly available for more than 1500 nearby galaxies, whereas obtaining interferometric observations of the gas at equivalent resolution remains prohibitively expensive. Therefore, while speculative, the dust correction map $s2$ is potentially a very powerful alternative.

As discussed in detail in \citet{2015ApJS..219....5Q}, the  $s2$ map contains all the non-stellar emission present at 3.6\,$\mu m$, which is essentially a diffuse mixture of PAH and continuum dust emission.
Therefore, it indirectly traces star formation (by hot dust and PAH emission), which is expected to correlate to some degree with the presence of molecular gas. In M51, we can test how good this correlation is by directly dividing the $s2$ map by the CO map from PAWS.
The ratio of $s2$ to CO is fairly uniform in the spiral arms and star-forming ring (0.009--0.033), while in the area of the bar it suddenly rises to 0.1--0.3, which means that it is a factor of 10 higher. This large difference provides a first clue that the dust map $s2$ may not be a reliable tracer of molecular gas in the central region of M51.

Figure\,\ref{fig:s2} confirms that the discrepancy between the torque profiles weighted by CO (PAWS) and by the map $s2$ is especially large towards the area of the nuclear bar. However, even without a one-to-one agreement, the \referee{correspondence} with HI is considerably better (the resulting torque profiles are good to about a factor of $\lesssim 2$). This suggests that the dust map $s2$ provided by the ICA separation can be used in a first-order approximation as a tracer of the ISM to weight the torques with, but this needs to be done with caution. In areas where the interstellar radiation field is strong (like the nuclear bar of M51), the $s2$ map can display significant flux even if CO emission is minimal, which could account for the differences we see between CO and $s2$ in the central region of M51. Additionally, we recommend using $s2$ as a tracer of the gas distribution only at low resolution; at high resolution is it less reliable because spatial offsets between tracers of the gas and star formation start to become evident.

\section{Dominant sources of uncertainty} 
\label{Sec:uncert}

We note that some sources of uncertainty have been considered previously
\citep[most notably, by][]{2009ApJ...692.1623H}, but many have not, including uncertainties intrinsic to the determination of the stellar mass distribution.  
We start by summarising the main sources of uncertainty involved in the torque and gas flow calculation and then consider their relative importance.

\begin{enumerate}

\item Projected stellar mass distribution:
   \subitem -- Importance of ICA dust correction: bias due to dust emission in 3.6\,$\mu m$ image.
   \subitem -- Effectiveness of ICA dust correction: imperfections in the stellar map introduced by ICA.
   \subitem -- Variations in stellar M/L.

\item 3D mass distribution and torques:
   \subitem -- Deprojection: 
   Accuracy in PA, inclination.
   \subitem -- Vertical disc structure: implications of assuming an isothermal disc and variations in disc scaleheight.
   \subitem -- Centring of stellar mass map.
   \subitem -- Centring of gas map.

\item Choice of ISM tracer:
   \subitem -- Implications of neglecting HI in the central region.
   \subitem -- Large-scale smooth component (short spacings)
   \subitem -- Variations in $X_\mathrm{CO}$
   \subitem -- Effect of resolution (1$''$, 3$''$, 6$''$)
   \subitem -- Use of dust correction map to estimate radial gas flows
\end{enumerate}


In the next subsections and in Appendix\,\ref{sec:appendix}, we present a number of tests to quantitatively assess the \corr{impact} of these uncertainties on torques and inflow estimates. The results of these tests are summarised
 in Table\,\ref{table:uncert}. We used the torques derived from our best ICA dust-corrected stellar mass map and the molecular gas distribution traced by PAWS as our fiducial result. For the various tests, we computed the relative difference with respect to the fiducial result, and the weighted mean is presented as an estimate of the relative uncertainty. We used the weighted average to avoid divergencies associated with values close to zero, which would unrealistically bias the mean difference; for this reason, the corresponding fiducial profiles, $\tau(r)$ or $d^2M/dr\,dt$ were used to provide those radial weights.

As shown by \citet{2014ApJ...788..144M} and \citet{2014ApJ...797...55N}, the 3.6\,$\mu m$ band minimises to a large extent the uncertainties in mass estimation, in the sense that it minimises extinction problems and allows for a single M/L to be applicable \citep[with an uncertainty of 0.1\,dex, smaller than the 0.2-0.3\,dex associated with prescriptions based on optical colours, which are often used to calibrate other NIR bands, e.g.][]{2009MNRAS.400.1181Z}. However, mass estimates based on 3.6\,$\mu m$ require proper accounting for dust emission, which can contribute as much as $\sim 30\%$ of the flux \textit{\textup{globally}} in star-forming spirals.
We assess the relevance of that critical correction for dust emission when the stellar mass distribution is calculated using a 3.6\,$\mu m$ image in Sect.\,\ref{Sec:importanceICAcorrection}.
This is the most critical uncertainty that had not been considered before, along with the resolution of the gas tracer, which we study in Sect.\,\ref{Sec:resolution}. The former has an impact of the order of $\sim 100\%$ in the case of M51 (a factor of 2), while the latter acts to `wash out' the signatures of the torque profiles, leading to much flatter profiles, and resulting in a difference of also $\sim 100\%$ when the resolution is degraded from 1$''$ to 13$''$.

The uncertainties on the position angle of the disc ($PA$), its inclination ($i$), and the correct centring of the stellar mass map and the gas map are likewise important, as has been shown by \citet{2009ApJ...692.1623H}. In Appendix\,\ref{sec:appendix} we confirm that these effects are significant for the configuration of M51 and that they show a radial dependence that causes the inner bins to be more strongly subject to centring errors and bins at large radii more uncertain because of errors in the deprojection parameters ($PA$, $i$). Fortunately, we have access to precise measurements of the $i$ and $PA$ \citep{2014ApJ...784....4C}, and the astrometry of our images (stellar mass map and gas map) was carefully calibrated to $<1''$ \citep{2013ApJ...779...42S}. The relevance of short spacing corrections on the interferometric gas maps was also stressed by \citet[][see their Fig.\,20]{2009A&A...496...85G} and \citet[][see their Fig.\,12]{2011A&A...529A..45V}. Errors can be as high as a factor of 2 for some radii in the case of missing short spacings; this study, however, did not suffer from this problem because our PAWS map includes short spacing information that ensures the recovery of all the flux \citep[see][]{2013ApJ...779...43P}.

For completeness, we also show in Appendix\,\ref{sec:appendix} that other possible uncertainties listed in Table\,\ref{table:uncert}, such as the vertical disc structure (function and scaleheight) or the radial variation of M/L, are clearly subdominant, with an \corr{impact} $\lesssim 10\%$. In principle, a radially varying M/L could have an \corr{impact} of as much as $\sim 20-30\%$ (as listed in Table\,\ref{table:uncert}), but this is an upper limit for galaxies with strong abundance gradients; for M51, the measured metallicity gradient is very shallow \citep[$0.02 \pm 0.01$\,dex\,kpc$^{-1}$][]{2004ApJ...615..228B}.
We can therefore safely conclude that M/L variations within its disc will not pose a large problem for measuring torques and gas flows.
 
 Finally, we also demonstrate in Appendix\,\ref{Sec:oversubtr} that the uncertainty introduced by oversubtractions in the stellar mass map (a drawback of assuming only two sources with ICA) is modest. 
In fact, these oversubtractions are not a concern in the area of the nuclear bar, while the ICA correction clearly is important in that region because a considerable amount of dust emission needs to be accounted for. \refereetwo{In Appendix\,\ref{sec:F190N} we double-check the validity of the ICA-corrected stellar mass map by comparing it with the gravitational potential obtained from an independent band (HST 1.9\,$\mu m$).}
 
\subsection{Importance of ICA dust correction}
\label{Sec:importanceICAcorrection}

\citet{2015ApJS..219....5Q} showed that the difference between the uncorrected 3.6\,$\mu m$ images and our ICA dust-corrected stellar mass maps is not only caused by hot dust in HII regions, but also by diffuse dust heated by the interstellar radiation field. Specifically, we identified considerable dust emission with ICA for M51 in the area of the nuclear bar (within $R < 22'' \sim 1$\,kpc). As much as 26\% of the 3.6\,$\mu m$ light is due to dust emission within the central $R<22''$ covered by the nuclear bar (see Fig.\,\ref{fig:depro-torques}), and the ratio of dust emission (as identified with ICA) to CO (via PAWS $1''$) increases from an average of $\sim 0.02$ in the PAWS field of view to $\sim 0.2$ in the area of the nuclear bar (a factor of 10).

The average change in the torque profile is a factor of 2 when the dust correction is not taken into account (100\,\%, over the whole extent of the profile; 70\,\% if we only consider the area of the nuclear bar, within $R < 22''$). Moreover, while the calculation based on the original, uncorrected 3.6\,$\mu m$ image led to positive profile values in the inner region ($r < 7'' = 250$\,pc; see Fig.\,\ref{fig:profiles}), suggesting a radial outflow that would impede the flow of gas to the AGN, the ICA dust-corrected stellar mass map implies molecular gas inflow 
\referee{down to our resolution limit},
leading to a qualitatively different interpretation. This stresses the importance of correcting 3.6\,$\mu m$ images for dust emission when the images are used to trace the distribution of stellar mass and gravitational torques are to be calculated.

\subsection{Effect of the gas tracer resolution (1$''$, 3$''$, 6$''$, 13$''$)}
\label{Sec:resolution}

\begin{figure*}[t]
\begin{center}
\includegraphics[trim=0 420 0 0,clip,width=1.0\textwidth]{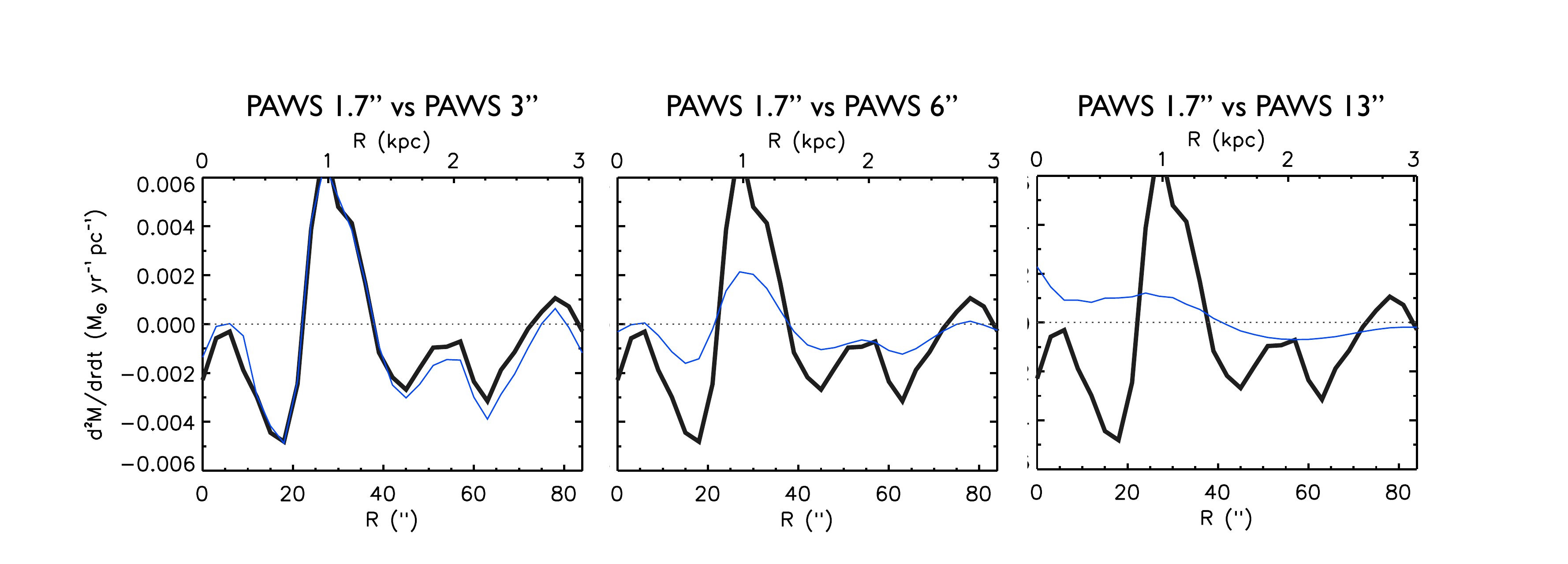}
\end{center}
\caption{Change in the torques that is due to the use of a gas map with different spatial resolutions. The thick blue line corresponds to the PAWS CO(1-0) intensity map at the resolution of the stellar mass map (1.7$''$), the blue line in the profiles from left to right correspond to the same PAWS map smoothed to 3$''$, 6$''$ and 13$''$.}
\label{fig:PAWSresolution}
\end{figure*}

In Fig.\,\ref{fig:PAWSresolution} we analyse the \corr{impact} of spatial resolution on the profiles. Changing the resolution of the gas tracer will have an effect on the $\tau(r)$ torque profiles because the torques are weighted by the gas distribution when computing the azimuthal averages. However, the effect becomes even stronger
when measuring gas flows because in this case the gas column density is directly part of the equation; this is why we show the profile $d^2M\,/\,dr\,dt$ instead of $\tau(r)$ in
Fig. 7.

To determine the resolution effect, we compared the different versions of the PAWS map, $1''=40$\,pc, $3''=120$\,pc and $6''=240$\,pc (tapered), and $13''=520$\,pc (smoothed), corrected for the nuclear outflow component as described in Sect.\,\ref{Sec:outflow}. For reference, we recall that the nuclear bar has a radius of $R_\mathrm{bar} \approx 20''$ and the spirals arms traced by CO have a typical width of around $\sim 10''$.
We verified that the results do not change if (instead of tapering) we smooth the PAWS moment-0 map with the corresponding Gaussian kernel (FWHM=$3''$, $6''$, $13''$). We note that 6$''$ and 13$''$ are the angular resolutions provided by the THINGS HI and HERACLES CO surveys. For consistency in terms of resolution, we also smoothed our stellar mass map to match these resolutions following an analogous procedure.

Degrading the resolution from $1.7''$ to $3''$ is associated with only a modest change of the gas flow rates (26\%). However, the average change is large when we move to 6$''$ (64\%) and becomes even more so at a resolution of $13''$ (90\%). This stresses that a map of the gas distribution is necessary that probes the adequate spatial scales; if the resolution of the gas tracer is too low ($\gtrsim 200$\,pc), the torque profiles are artificially smoothed and important structure becomes lost.

\subsection{Typical final uncertainty}
\label{Sec:finaluncert}

As we have seen, the factors that would most severely hamper the analysis of gravitational torques are the absence of dust correction in the 3.6\,$\mu m$ images and using a gas map with poor resolution; these two are not a concern for our analysis. The next most significant sources of error are the uncertainty in the deprojection parameters ($PA$, $i$), which can introduce an error of as much as $\sim 30-50\%$. Similarly, the incorrect centring of the NIR image can have a very strong effect on the torques, but  we confirmed our astrometry to an accuracy better than $1''$, therefore the error should typically be smaller than $\sim 20\%$ (the innermost bin is a special case discussed below). The centring of the gas image is even less problematic because the associated uncertainty is smaller and our astrometry is even more accurate.

As shown in the appendix (Fig.\,\ref{fig:radialerrors}), the uncertainty introduced by an error in the determination of deprojection parameters ($PA$, $i$) has a clear radial dependence, becoming particularly important in the outer regions of the galaxy ($\sim 100\%$ at a radius of 3\,kpc). In the inner 1\,kpc, only the uncertainty due to errors in $i$ could be significant, $\sim 40\%$ (the uncertainty due to errors in $PA$ is below $\sim 10\%$). Conversely, for the centring of the NIR image, the radial dependence of the uncertainty is reversed, with larger errors for smaller radii (up to $\sim 50\%$ in the first bin).
As a consequence of this, the innermost bin is subject to considerably high uncertainties, which add to potential complications associated with the outflow component that we subtracted in Sect.\,\ref{Sec:outflow}. 
\refereetwo{Appendix\,\ref{sec:F190N} also suggests divergencies in the inner $R \lesssim 5''$ when using an independent band to compute the gravitational potential (HST 1.9\,$\mu m$).}
Therefore, the innermost bin has to be treated with caution, and we have shaded it in Fig.\,\ref{fig:profiles}.

\begin{table*}[t!]
\begin{center}
\caption[h!]{Effect of the uncertainties in different input parameters on gravitational torques and inflow rates for M51.}
\begin{tabular}{lcccc}
\hline\hline 
Uncertainty due to...           &       $\Delta t$ (\%) &       $\Delta t$ for $R<22''$  (\%)   &       $\Delta (d^2M/dr\,dt)$  (\%)    &       $\Delta (d^2M/dr\,dt)$ for $R<22''$  (\%)       \\
\hline 
Uncorrected, original 3.6\,$\mu m$ image        &       101.3   &       70.7    &       93.2    &       73.0    \\
Inclination ($\pm 5^o$) &       51.7    &       39.3    &       45.3    &       50.1    \\
Position angle ($\pm 3^o$)      &       34.6    &       6.7     &       21.0    &       19.0    \\
Radially varying M/L    &       21.4    &       27.8    &       20.5    &       27.5    \\
Shift of stellar mass map (0.75$''$)    &       20.3    &       71.2    &       34.3    &       80.6    \\
Shift of gas image (0.75$''$)   &       12.6    &       27.2    &       17.5    &       33.7    \\
Stellar disc scale-height ($\pm 20 \%$)  &       5.1     &       10.8    &       6.4     &       11.0    \\
Stellar disc vertical function (sech$^2$ vs sech)        &      2.4     &       5.4     &       3.1     &       5.6     \\
        \hline                                                          
PAWS: 1.7$''$ vs 3$''$  &       38.3    &       32.6    &       26.1    &       22.2    \\
PAWS: 1.7$''$ vs 6$''$  &       43.9    &       32.0    &       64.0    &       72.0    \\
PAWS: 1.7$''$ vs 13$''$ &       84.5    &       111.2   &       90.1    &       103.3   \\
PAWS 6$''$ vs THINGS 6$''$      &       79.7    &       65.6    &       85.8    &       85.1    \\
PAWS 13$''$ vs HERACLES 13$''$  &       32.2    &       40.0    &       41.0    &       34.0    \\
 \hline
\end{tabular}
\label{table:uncert}
\end{center}
\end{table*}

\section{Discussion}
\label{Sec:Discussion}

 
The feedback from AGN plays a critical role in reconciling observations with cosmological simulations of galaxy evolution, prevents galaxies from over-growing, and it might hold the key to the tight scaling relations observed between the mass of supermassive black holes and their host galaxies. However, this important process is ultimately regulated by the availability of fuel in the nucleus, and consequently, the transport of gas across the galaxy is a crucial process.  The gas might be accreted and continuously replenished from the circumgalactic medium, but in the absence of effective mechanisms to transport it to the centre, the fuel will never reach the AGN.

In this context, we have studied the gas transport in the disc of M51, making use of our stellar mass map (based on 3.6\,$\mu m$ imaging and corrected for dust emission with ICA). We have shown that the use of a proper stellar mass map is critical, as is the use of a gas tracer at sufficiently high spatial resolution (see Table\,\ref{table:uncert}, which summarises the tests that we have presented in Sect.\,\ref{Sec:uncert}). Even if the analysis of all the possible sources of uncertainty might sound pessimistic, the radical improvement of instrumental capabilities and the advancement of techniques allow us to enter a regime where this can be overcome. With ALMA and NOEMA, maps of the molecular gas distribution at sufficiently high spatial resolution are expected
to be easily achievable. 
Moreover, in the mid-term future, the advent of JWST should make it possible to obtain proper stellar mass maps using our technique or others with even higher resolutions and for more galaxies.

\subsection{Inflow rates}
\label{Sec:inflow}

The main result of the paper is that torques imply consistent molecular gas inflow from $R=22'' \sim 1$\,kpc down to \referee{our resolution limit when the ICA-corrected
stellar mass map of M51 is used } (but we note that the inner bin, $R<3''$, is subject to large uncertainties; \refereetwo{see also Appendix\,\ref{sec:F190N}}). Torques become positive after $R=22''$, implying radial outflow, and become negative again at a radius of $R=37''$. This agrees with the presence of a nuclear molecular ring, which could have formed through the accumulation of molecular gas as a consequence of the flow pattern that we have discussed, and which would imply a dynamical barrier for gas transport. There must be other mechanisms (viscous torques, instabilities, etc.) that allow the gas to overcome this barrier and access the area of the nuclear bar, but they probably operate over longer timescales, and that is the reason for the formation
of the overdense molecular gas in the shape of a ring.


We find that as much as $\sim 5\,M_\odot /yr$ of molecular gas are involved in net inflow motions in the central 1\,kpc of M51. In its way towards the centre, part of this amount of gas will form stars and never reach the nucleus; for reference, the integrated star formation rate within the central $20''$ is around $0.5\,M_\odot /yr$ \citep{2007ApJ...671..333K}. At the smallest scales accessible to us ($R \sim 3'' \sim 110$\,pc), the inflow rate is $\sim 1\,M_\odot /yr$. Even if we ignore this innermost bin because of the increasingly large uncertainties, the second and third bins imply virtually the same inflow value, which suggests that the result is \corr{robust}. These net inflow rates should be regarded as a lower limit, since we did not attempt to account for viscous torques or other dissipative effects (cloud-cloud collisions, shocks, etc.). However, these effects usually operate on much longer timescales, therefore the global value is expected to be close to the value that we have derived using torques. After gas has reached a radius of 100\,pc (the outer edge of our innermost bin), dynamical friction effects are expected to quickly bring the gas close enough to the SMBH (Combes 2002).

This observationally estimated inflow rate is an important number, for instance, to compare with the rates obtained using simulations of M51 or other grand-design spiral galaxies. In particular, the simulation of the M51 system presented in \citet{2010MNRAS.403..625D} allows us to measure the flow of molecular gas and compare this
to our observations. As we show in Appendix\,\ref{sec:appendixsimu}, the interacting system evolves towards a more centrally concentrated molecular gas distribution, while an analogous model in isolation does not show such a trend. The change in H$_2$ surface density in the simulation implies molecular gas inflow at a rate of $\sim 2\,M_\odot /yr$, which is similar to our observational findings. 
From a future perspective, gathering similar inflow rates for more galaxies will permit accessing valuable statistics, which might help to impose constraints on cosmological simulations.

\subsection{Dynamical resonances}
\label{Sec:dynamres}

Our analysis of the new gravitational torque profiles for M51 has led us to suggest that the galaxy has two main corotations: one associated with the nuclear bar, approximately at CR$_\mathrm{bar} \sim 20''$, with a pattern speed of 
$\Omega_P^\mathrm{bar} =(185 \pm 15)$\,km\,s$^{-1}$\,kpc$^{-1}$, and one for the spiral pattern near CR$_\mathrm{sp} \sim 100''$, implying a pattern speed of $\Omega_P^\mathrm{spiral} =(53 \pm 5)$\,km\,s$^{-1}$\,kpc$^{-1}$. The other resonances associated with these two corotations are summarised in Table\,\ref{table:reson}.

The corotation of the spiral occurs much farther out than the crossing near $R=60''$ suggested by \citet{2013ApJ...779...45M} as a potential location for the spiral corotation, which should be considered superceded by our more \corr{robust} determination at $100''$. \citet{2013ApJ...779...45M} also used the technique of constructing radial torque profiles to assess the positions of resonances, but there are a number of factors that account for the difference between both results. First, the torque map of \citet{2013ApJ...779...45M} was compromised by remaining non-stellar emission, which only our improved strategy has been able to identify \citep[through an iterative implementation of ICA, see][]{2015ApJS..219....5Q}. The difference in the 
dust fraction identified with ICA is from 8\% \citep{2013ApJ...779...45M} to 34\% \citep{2015ApJS..219....5Q}. We note that, specifically, the region from $40-80''$ is characterised by a high incidence of HII regions along the spiral arms (hence the large change of the torque profile in that area given our second iteration of ICA, which is designed to better accommodate for the multi-source mix of hot dust and PAHs).
 
 Secondly, the new tests that we performed here have shown that the torque profiles based on PAWS are not meaningful outside $85''$; even if PAWS provides information out to a radius of $R \sim 120''$, the coverage is not uniform in the range $R \sim 85 - 120''$ because of the rectangular field of view of the survey. Even at a radius of $R \sim 90''$ (only $5''$ beyond the last possible circular aperture) the difference between the two profiles can be as much as a factor 2, and this only deteriorates as the radius increases and the sampling is more incomplete. 
We therefore opted to limit our measurements with the PAWS data to within $R<85''$ and extracted measurements of torques at larger radii with THINGS, which covers a full $0-2\pi$ radians out to $R \sim 20$\,kpc, albeit with lower spatial resolution.

We also used HERACLES, which traces molecular gas (as opposed to THINGS, tracing atomic gas) to confirm the analysis out to larger radii; this should allow us to avoid the biases that we just discussed that are due to the non-circular shape of the PAWS field of view. 
In principle, there might be important differences between THINGS and HERACLES, not only because of the different spatial resolution, but also because HI and CO emission are not exactly co-spatial \citep[see][]{2013ApJ...779...42S}, and they also have significant kinematic differences \citep{2014ApJ...784....4C}. However, despite tracing different phases of the ISM and having different resolutions, the bottom panel of Fig.\,\ref{fig:angfreq} shows that the corresponding torque profiles are reasonably similar, and they have major negative--positive zero-crossings at similar positions ($R \sim 20''$, the CR of the nuclear bar; $R \sim 100''$, the proposed CR of the spiral). The qualitative agreement between both profiles (and the analysis of $s1/s3$ terms) reinforces the scenario that we proposed here.

Additionally, our estimates for the two (bar and spiral) corotation radii in M51 are consistent with the values independently estimated by \citet{2012arXiv1203.5334Z} ($R_\mathrm{CR1}=25''$ and $R_\mathrm{CR2}=110''$). The corotation of the bar also agrees with \citet{2014ApJ...784....4C}, who suggested a radius of $R_\mathrm{CR}^\mathrm{bar}=20''$ that
was confirmed visually and analytically with a clear change from $m-1$ to $m+1$ modes in the residual velocity field \citep{1993ApJ...414..487C}. Their proposed resonances for the spiral, however, mostly relied on the torque result from \citet{2013ApJ...779...45M}, which suffered from the problems mentioned above. 
\referee{It is interesting to note that their proposed $m=3$ mode could play the role of dynamically coupling the bar with the spiral structure, explaining the position of the inner pseudo-ring around $R \sim 20''$.}

From a broader perspective, these results raise interesting questions about the nature of spiral structure. The interaction with the companion NGC\,5195 has been suggested to drive dynamical changes over relatively short timescales ($\rm \sim 100\,Myr -- 1 Gyr$), which might contradict the na\"{i}ve picture of a long-lasting density-wave spiral. For example, \citet{2010MNRAS.403..625D} did not find a single global pattern speed for M51 in their numerical simulation and instead suggested a picture in which the spiral arms are the result of tidally induced local kinematic density patterns that wind up (i.e. the pattern speed varies with radius). In their simulation, corotation can only exist beyond a radius of 7\,kpc (probably around 10\,kpc, close to the satellite NGC\,5195). Salo and Laurikainen (2000) also used numerical simulations to show that the close passage of the companion removes any previous hints of spiral structure, suggesting that the spiral appearance that we see today is relatively new and might not persist very long. 

In this context, we need to be very cautious when interpreting the observational torque profiles in terms of resonances because ILR, OLR and other resonances can also have an imprint in the form of a change of sign.  The fact that the THINGS profile only shows two crossings from negative to positive is reassuring, but it could be that the second one is not corotation, but rather an inner resonance of the spiral that might have corotation much farther out, even outside our fields of view. This possibility cannot be ruled out with our data.

\section{Summary and conclusions}
\label{Sec:conclusions}

We have used the new $3.6\,\mu m$ dust-corrected stellar mass map from \citet{2015ApJS..219....5Q} to construct a gravitational torque map for M51, which, in conjunction with the PAWS dataset, has allowed us to estimate radial gas flows in this galaxy.

\begin{enumerate}

  \item There is plenty of molecular gas inflow in the  central area of M51: $\sim 5$\,M$_\odot$/yr in the central $20 '' \sim 1$\,kpc, $\sim 1$\,M$_\odot$/yr as we approach \referee{our resolution limit (60\,pc} from the nucleus). This inflow is driven by the nuclear bar.
  
  \item Torques suggest a corotation of the bar at $R \sim 20''$ and corotation of the spiral at $R \sim 100''$.
  
  \item We demonstrated the importance of correcting 3.6\,$\mu m$ images for dust emission when estimating torques. For these purposes, differences can reach a factor of 2. In the case of M51, even the qualitative result of whether or not there is inflow in the central kpc changes as a consequence of the dust correction in the central area.
  
  \item We also assessed the difference due to varying M/L and uncertainties in inclination, PA, etc., showing that the effect of M/L is modest, and confirming the results from \citet{2009ApJ...691.1168H} for the other parameters (largest uncertainty due to centring of NIR image and deprojection parameters).
  
  \item We considered different ISM tracers for the azimuthal weighing of the torques. We showed that the spatial resolution of the molecular gas is critical (better than $\sim 100$\,pc needed), and the contaminant map $s2$ can be used at most as a first-order approximation to identify the presence of dynamical resonances and gas flows.

\end{enumerate}

In conclusion, this study provides the first estimation of radial gas flows in M51, with an inflow rate that can be checked against simulations and can be compared with outflow rates. We also carefully analysed the limitations in the calculation of gravitational torques, which should be useful when planning similar studies in other galaxies; this will be especially relevant as ALMA and NOEMA become fully operative.
In particular, we stressed the importance of using proper stellar mass maps when estimating radial gas flows in nearby galaxies, in addition to the strong limitation imposed by the resolution of the available molecular gas observations. Thanks to the efforts of S$^4$G and the public data release of accurate stellar mass maps through IRSA for more than 1500 nearby galaxies, with the spatial resolutions achieved with ALMA and NOEMA, these estimations should be feasible in the opening ALMA and NOEMA era.

\small  
%
\begin{acknowledgements}   
\refereetwo{The authors would like to thank the anonymous referee for a helpful report, as well as Daniela Calzetti, Nick Z. Scoville and Mari Polletta for making the HST/F190N mosaic available to us.}
We also
 appreciate valuable comments from Fran\c{c}oise Combes and Sebastian Haan. 
We acknowledge financial support to the DAGAL network from the People Programme (Marie Curie Actions) of the European Union's Seventh Framework Programme FP7/2007- 2013/ under REA grant agreement number PITN-GA-2011-289313. M.Q. acknowledges the International Max Planck Research School for Astronomy and Cosmic Physics at the University of Heidelberg (IMPRS-HD). S.G.B. thanks support from Spanish grant AYA2012-32295. \refereetwo{J.P.
 acknowledges support from the CNRS programme ``Physique et Chimie
 du Milieu Interstellaire'' (PCMI). M.Q., S.E.M., D.C. and A.H. acknowledge funding from the Deutsche
Forschungsgemeinschaft (DFG) via grants SCHI~536/7-2,
} SCHI~536/5-1, and SCHI~536/7-1 as part of the priority program SPP~1573 ``ISM-SPP: Physics of the Interstellar Medium.''
\end{acknowledgements}

\appendix
\section{Sources of uncertainty involved in torque and gas flow estimations}
\label{sec:appendix}

\subsection{Imperfections introduced by the ICA dust correction}
\label{Sec:oversubtr}

\begin{figure*}[thp]
\begin{center}
\includegraphics[trim=-35 400 12 80,clip,width=1.0\textwidth]{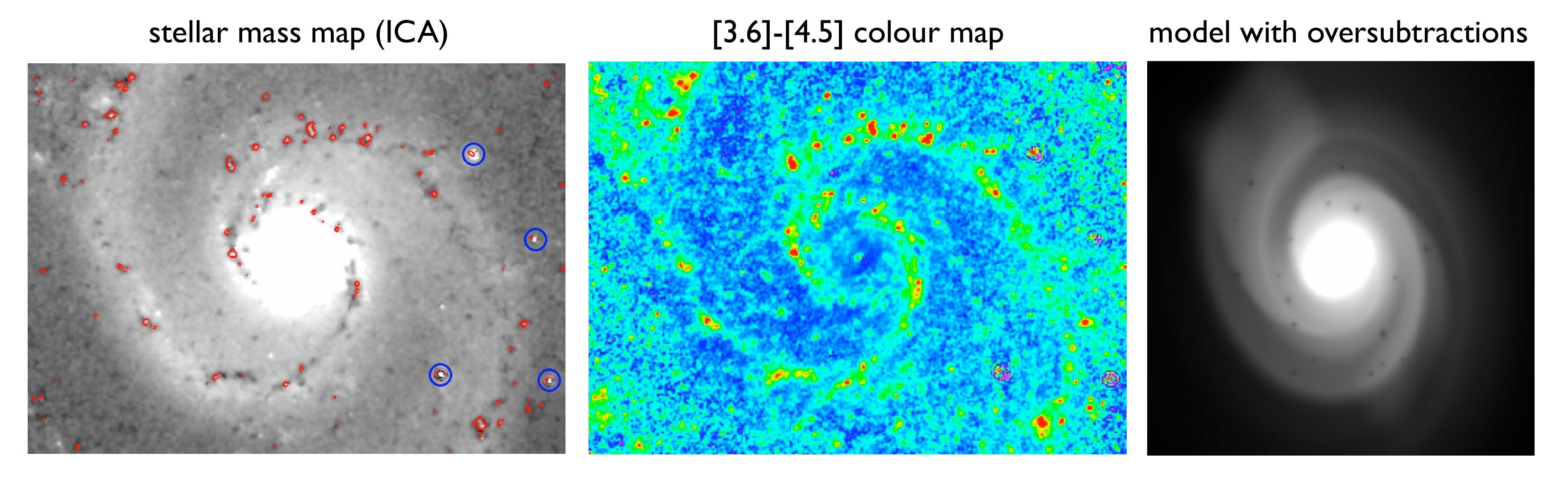}
\end{center}
\caption{\textit{Left:} stellar mass map of M51, shown on a scale that emphasises the oversubtractions associated with the ICA dust correction; the red contours are the regions of M51 where the [3.6]-[4.5] colour is very red, basically corresponding to HII regions ([3.6]-[4.5]>0.2). The blue circles mark the positions of field stars, with the clear signature of concentric circles in the colour map due to a saturated PSF.
\textit{Middle:} [3.6]-[4.5] colour map of M51; compare the lack of very red regions in the central $R < 22'' \sim 1$\,kpc with their presence in the spiral arms.
\textit{Right:} one of the 40 random realisations of the model with oversubtractions as described in the text (Sect.\,\ref{Sec:oversubtr}), which we use to assess the effect of these imperfections on gravitational torques.}
\label{fig:oversubtractions}
\end{figure*}

\begin{figure*}[th]
\begin{center}
\includegraphics[trim=0 420 -10 0,clip,,width=1.0\textwidth]{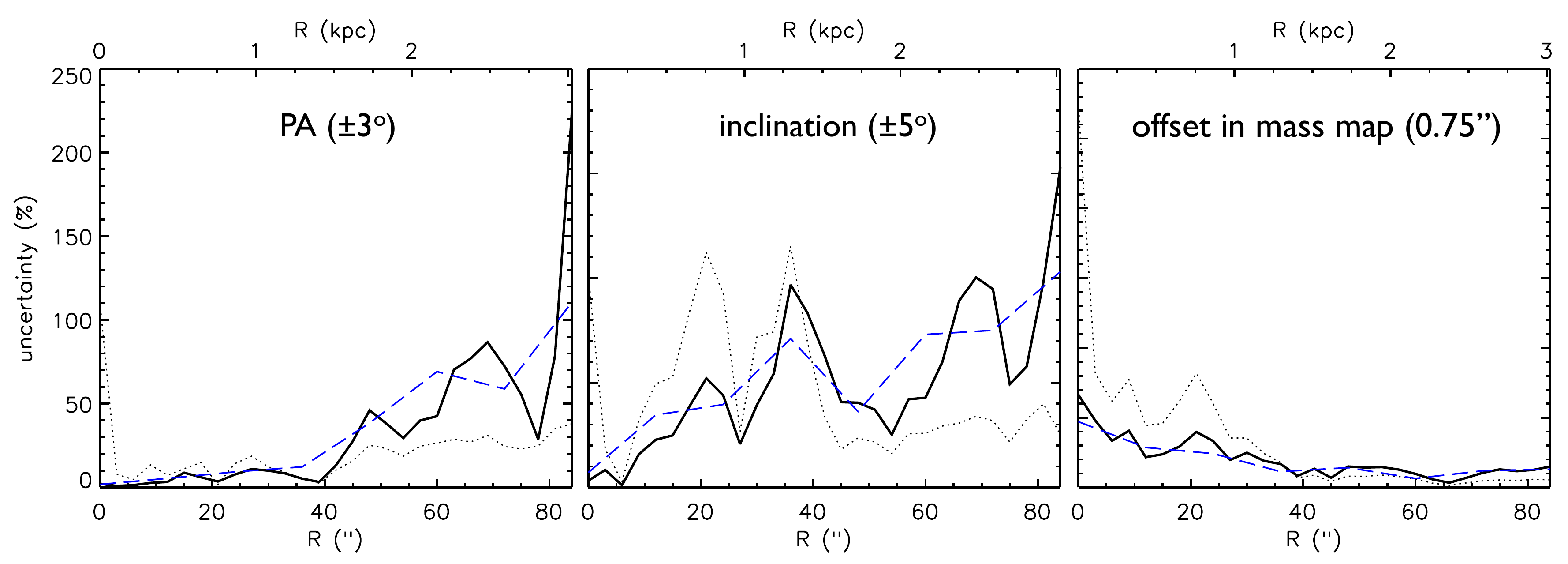}
\end{center}
\caption{Radial dependence of uncertainties associated with PA ($\pm 3^\circ$), inclination ($\pm 5^\circ$) and an offset of 1px in the centring of the NIR image. The continuous black line shows the effect on torques, the dotted line corresponds to the effect on the local flow rates, and the dashed blue line is the average tendency (effect on torques averaged over of regions of $\sim 400$\,pc).}
\label{fig:radialerrors}
\end{figure*}

\begin{figure*}[tp]
\begin{center}
\includegraphics[trim=40 240 120 0,clip,width=1.0\textwidth]{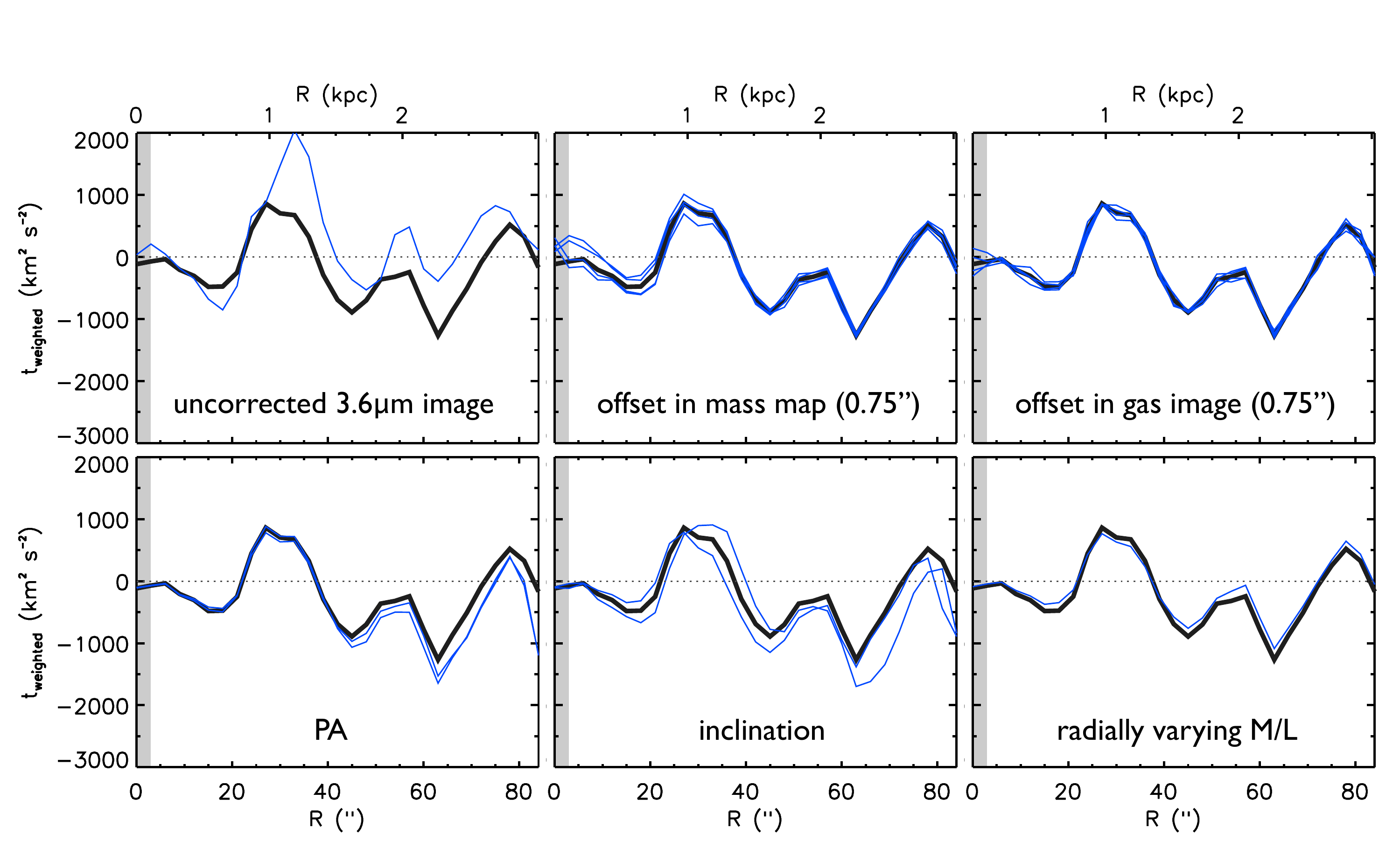}
\end{center}
\caption{Change in the torque profiles due to a number of systematic uncertainties. The thick black line is our fiducial profile, based on the ICA dust-corrected 3.6\,$\mu m$ image. From top left to bottom right, the thin blue lines are the profiles based on the uncorrected 3.6\,$\mu m$ image (i.e. difference due to dust emission at 3.6\,$\mu m$); an offset in the centring of the stellar mass map of $0.75''$; an offset in the centring of the gas map (CO traced by PAWS) of $0.75''$; a variation in the position angle of the disc corresponding to the extremes of the error bars given by \citet{2014ApJ...784....4C}; a variation in the inclination of the disc corresponding to the extremes of the error bars given by \citet{2014ApJ...784....4C}; and the
difference in mass distribution associated with the maximum possible radially varying M/L.}
\label{fig:uncertainties}
\end{figure*}

As commented in \citet{2012ApJ...744...17M} and \citet{2015ApJS..219....5Q}, the choice of only two components for the ICA separation can introduce some imperfections in the output stellar mass map. These come mostly in the form of oversubtractions corresponding to HII regions, where the [3.6]-[4.5] colour of the hot dust is considerably redder than the average dust colour found by ICA; when the 3.6\,$\mu m$ image is separated into two components using the scaling given by the stellar and dust colours solved for by ICA, the regions of intrinsically redder colour than average will be overestimated in terms of dust, leading to an oversubtraction in the stellar component. The problem can be alleviated by a sophisticated interpolation technique or by masking these regions. However, here we show that for the purpose of estimating gas flows based on the corresponding gravitational torque map, the \corr{impact} of leaving these imperfections in the stellar mass map is not very large. Moreover, as becomes evident from Fig.\,\ref{fig:oversubtractions}, the oversubtractions are completely absent from the region of the bar, which means that they would not affect the inflow claim that we have made for the central $R<1$\,kpc.

An inspection of our M51 stellar mass map with attention to the areas of very red colours ([3.6]-[4.5]>0.2, which is similar to imposing a sufficiently high H$\alpha$ flux) shows that the `oversubtractions' have average sizes of $\sim$6\,px in diameter (minimum $\sim$3\,px, maximum $\sim$12\,px in diameter; i.e. $\sim 2-9'' \sim 80 - 320$\,pc; see Fig.\,\ref{fig:oversubtractions}). The contrast of these oversubtractions, measured as the minimum value in the centre of the oversubtraction divided by the average value in the immediate surroundings, is typically $\sim$0.7 (a contrast value of 1 would indicate no oversubtraction at all; the contrast of 0.7 means that the oversubtractions typically involve a $\sim$30\% depression with respect to the local flux); we have also confirmed that the standard deviation of the contrast from one region to another is small ($\lesssim$0.1). 

In an attempt to quantify the \corr{impact} of these oversubtractions, we used a smooth, multi-component Galfit model of M51 \citep{2010AJ....139.2097P} and imposed a random distribution of oversubtractions that scale to the local flux (one of these realisations is also illustrated in the right panel of Fig.\,\ref{fig:oversubtractions}). The imperfections were modelled as de Vaucouleurs profiles that were subtracted from the smooth model, with the central peak brightness of each oversubtraction being a fraction of the local flux (contrast of 0.7), and with a scale length that reflects the typical range of scale lengths covered by the oversubtractions in the ICA dust-corrected stellar mass map (we assumed a normal distribution for the sizes, with a mean of 6\,px and a standard deviation of 2\,px). We performed a battery of 40 such experiments, in which the oversubtractions are placed randomly in the disc, wherever the model shows significant flux ($>0.6$\,MJy/sr). This leads to a typical change in the torque profile of $\sim 10\%$ ($0.13 \pm 0.06$). We also checked that forcing the positions of the oversubtractions to `cluster' by allowing them only to be placed where the colour map [3.6]-[4.5]>0.2 (or, equivalently, very high H$\alpha$ flux) does not have a dramatic effect on the torque profiles: the final uncertainty in the torque profile increases, but is still below 25\%. We imposed a total of 20 oversubtractions, but we checked that doubling this amount (40 oversubtractions) also keeps the final errors below 20\%.

Therefore, the uncertainty introduced on the final torque profiles by the imperfections associated with the ICA correction is of the order of $\sim 10\%$ and is reasonably bound to be below 20\% even under the most unfavourable conditions. This is much smaller than the change implied by the ICA correction itself ($>80\%$), which shows that the benefit and importance of using the dust-corrected stellar mass map clearly outweighs the drawback associated with these oversubtractions.

\subsection{Variations in stellar M/L}

As identified by \citet{2004ApJS..152..175M}, the most significant changes in ages and metallicity within spiral galaxies are radial, with a fairly smooth variation as a function of radius. The change is such that the inner regions are older and more metal-rich, while the outer regions are younger and more metal-poor. Even if we assume that the radial change in M/L that is due to these age and metallicity gradients is the maximum variation allowed by the \citet{2014ApJ...788..144M} conversion value (from 0.48 at $R=0$ to 0.75 at $R=R_{max}$, spanning the $0.6 \pm 0.1$\,dex uncertainty), this introduces only a modest change in the final torque profile ($\sim 20\%$), as is also visible in the bottom
right panel of Fig.\,\ref{fig:uncertainties}.

We note that our assumption is very conservative because the age and metallicity gradients within a given galaxy are in general much smaller than the changes from galaxy to galaxy, for which the \citet{2014ApJ...788..144M} uncertainty accounts.
Moreover, the measured metallicity abundance gradient in M51 is very shallow \citep[$0.02 \pm 0.01$\,dex\,kpc$^{-1}$][]{2004ApJ...615..228B}.
 Therefore, we can safely conclude that M/L variations within the disc of a galaxy will not pose a large problem for measuring torques and gas flows, and even less so for M51.

\subsection{Deprojection: accuracy in PA and inclination}

Figure\,\ref{fig:uncertainties} shows the effect of varying the deprojection parameters, inclination, and PA of the disc within the range covered by the error bars of the observational measurements \citep[$i=22 \pm 5$, PA=$173 \pm 3$;][]{2014ApJ...784....4C}. Our underlying assumption is that the molecular gas disc traced by CO is coplanar with the stellar disc, which seems to be a justified assumption; this leads to a considerable reduction of the uncertainties in these parameters (derived from kinematics) with respect to what could be obtained from photometry. For the particular configuration of M51, the change implied by the uncertainty in inclination and PA is significant, $\sim 50\%$ and $\sim 35\%$, respectively, when the whole extent of the profiles is considered. We note, like \citet{2009ApJ...692.1623H} have pointed out, that there is a radial dependence of the resulting errors. Specifically, for the deprojection parameters (PA, $i$), the \corr{impact} of the uncertainties becomes larger for larger radii; in the area of the nuclear bar, the uncertainties decrease to 39\% and 7\% for $i$ and PA, respectively. However, the errors in PA and $i$ can become dominant at large radii ($> 2-3$\,kpc): this radial dependence of the errors is demonstrated in Fig.\,\ref{fig:radialerrors}.

\subsection{Vertical disc structure}

We assumed that the vertical structure of the stellar disc corresponds to an isothermal disc with constant scale height; this is confirmed by observations \citep{1989ApJ...337..163W, 1992AJ....103...41B}
 and expected theoretically if the disc is in equilibrium. The vertical distribution of mass can be described by the following function:

\begin{equation}
\rho_z(z)=\rho_0 \sech^2(z/h).
\end{equation}

 We also checked that using other functions, such as sech instead of sech$^2$, provides almost exactly the same answer; the changes are $\lesssim 10\%$ in any case. The scale height is estimated as $h \sim 1/12 H_{\mathrm{disc}}$ \citep{1989ApJ...337..163W, 1992AJ....103...41B}, for which $H_{\mathrm{disc}} = 100''$ from the Galfit photometric decomposition of M51 \citep{2015ApJS..219....4S}. We also considered possible variations in the disc scale-height: as long as the variation in the scale height is kept below 20\%, the effect is negligible (average change of 5\% on the torque profile). Even a very large change in the scale height, of the order of 50\%, would only have an effect of $\sim 30$\% on the torques.


\subsection{Centring of gas and stellar mass maps}
\label{AppCentring}

As has discussed by \citet{2009ApJ...692.1623H}, one of the most critical points limiting the accuracy of the torques is the proper centring of the NIR image. Figure\,\ref{fig:uncertainties} shows the difference introduced by an offset of $0.75''$ and demonstrates the clear radial dependence of the error: it decreases with distance. The implied difference in the central $22''$ (880\,pc) is 71\%, while it drops to only 20\% if we consider all radii out to $90''$ (3\,kpc). This is also illustrated by Fig.\,\ref{fig:radialerrors}, in which the average radial dependence of the uncertainties is made explicit. 

For the 3.6\,$\mu m$ image and the stellar mass map based on it, the astrometric accuracy is lower than $1''$, as inferred from a comparison with the shifted SINGS 3.6\,$\mu m$ image \citep{2013ApJ...779...42S} using foreground point sources as the matching reference. This is reassuring because even if offsets in the stellar mass map are one of the main sources of uncertainty, in our case this uncertainty should be constrained to be clearly below $1''$, which involves a maximum average effect of $\sim 20$\%. However, as a result of the radial dependence that we have already noted, the inner bins will be subject to a larger uncertainty. 

As also pointed out by \citet{2009ApJ...691.1168H}, the centring of the gas map (e.g. moment-0 map of molecular gas intensity distribution) is not as crucial as the centring of the near-infrared image. This is because an error in the gas centring will only slightly modify the weights of the azimuthal average, while an incorrect centring of the stellar mass map will result in artificial torques.
We confirm for the case of M51 that the uncertainty introduced by an offset of 1\,px ($0.75''$) in the centring of the gas map is $\sim 10\%$. Given the accuracy of the astrometry in the PAWS map ($< 1''$), this should not be a problem.

\subsection{\refereetwo{Confirming torques with 1.9\,$\mu m$ imaging}}
\label{sec:F190N}

\begin{figure*}[tp]
\begin{center}
\includegraphics[width=0.92\textwidth]{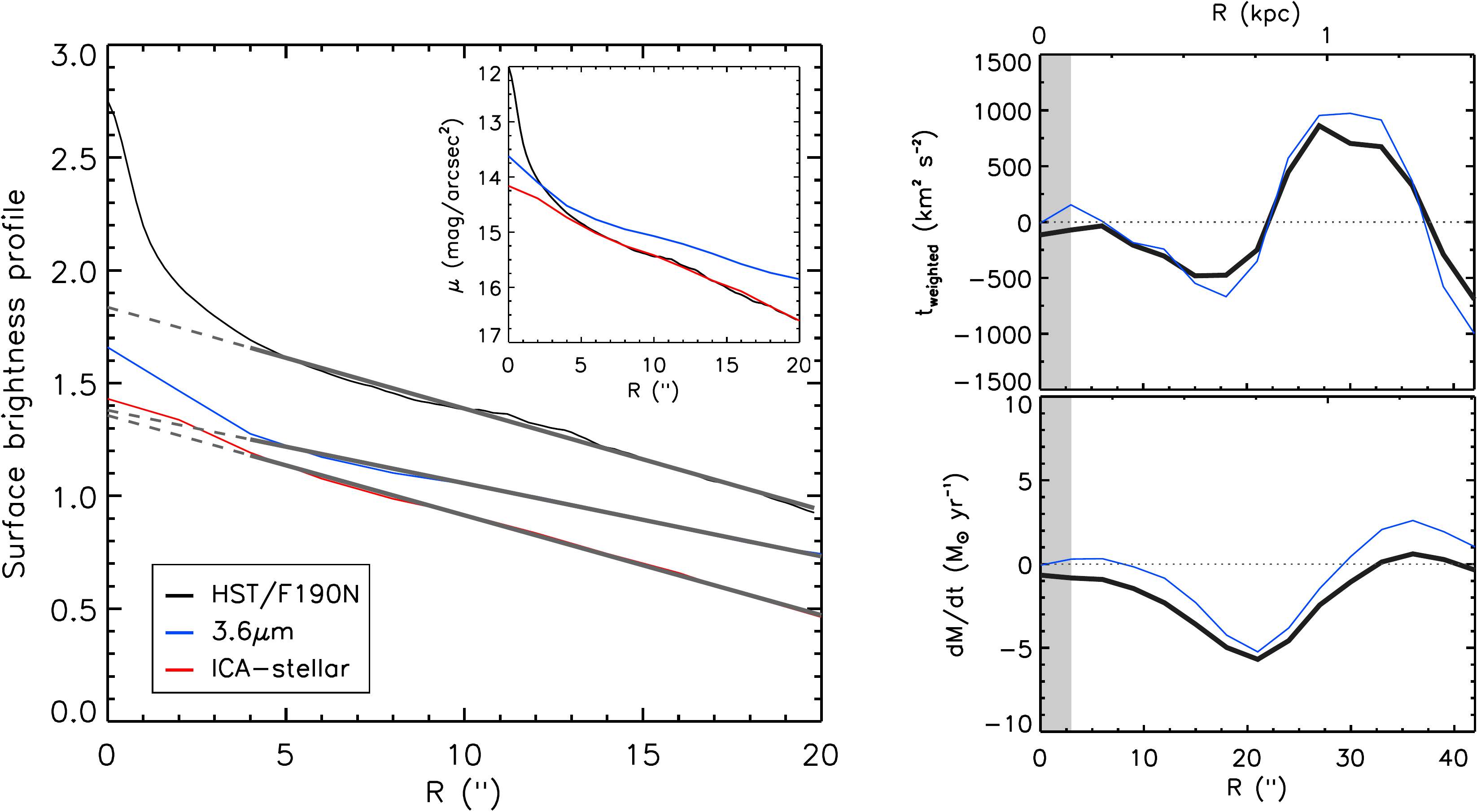}
\end{center}
\caption{\refereetwo{\textit{Left:} radial surface brightness profiles for the HST/F190N narrow-band image of M51, the  uncorrected 3.6\,$\mu m$, and the ICA-corrected 3.6\,$\mu m$ (stellar mass map) in units of $\log (F_\nu / \mathrm{[MJy/sr]})$. The linear fits in the range $5-20''$ highlight the good agreement with the expectation from a simple S\'ersic profile in this region and show that F190N and the stellar mass map have basically identical slopes. The small inset panel shows the equivalent surface brightness profiles in magnitude units (Vega): after accounting for the different zero points, the HST/F190N profile perfectly matches the levels of the stellar mass map. \textit{Right:} radial torque profile (top) and accumulated inflow rates (bottom) for the gravitational potential based on HST/F190N (blue), compared with our fiducial stellar mass map (black). The blue profile should be regarded as an (extreme) upper limit, in which the `bump' in the central region of HST/F190N is entirely attributed to old stars (which is highly unlikely, as argued in the text). The reason for divergencies out to $R \lesssim 5''$ in the top plot (torques) to propagate farther out in the bottom plot (inflow) is that the bottom profile displays \textit{\textup{cumulative}} flow rates, and an offset in the inner bin will shift the whole profile up or down.}}
\label{fig:F190N}
\end{figure*}

\refereetwo{As an additional sanity check, here we employ a 1.9\,$\mu m$ image to re-assess the gravitational potential and the inflow rates. We used the HST 1.9\,$\mu m$ continuum narrow-band image from \citet{2001AJ....122.3017S}, a 3$\times$3 mosaic of the central region of M51 (uniformly sampled only out to $R \sim 50''$) obtained with the F190N filter on the NICMOS\,3 camera (continuum next to the Pa\,$\alpha$ line); for details on data reduction see \citet{2001AJ....122.3017S}. The inflow rates and resonances are essentially identical as those presented in Sect.\,\ref{Sec:inflowrates}. Only in the central $R \lesssim 5''$ do the results using this HST image differ, but this has already been flagged as a problematic region associated with very high uncertainties.}

\refereetwo{Given the importance of image centring for torque computations (Appendix\,\ref{AppCentring}), we corrected the HST/F190N mosaic to an absolute astrometric frame of reference following the technique described in \citet{2013ApJ...779...42S}, Sect.\,7.3; this results in an astrometric offset of $1.3''$ ($\Delta x=+0.729''$, $\Delta y=-1.08''$). We assumed a constant mass-to-light ratio, $M/L_{1.9\mu m} = 0.2\,M_\odot/L_\odot$, chosen so that the stellar mass radially matches our stellar mass map in the range $5-20''$. In Fig.\,\ref{fig:F190N} (left) we show the radial surface brightness profile for HST/F190N, compared with the uncorrected 3.6\,$\mu m$, and the ICA-corrected 3.6\,$\mu m$ (stellar mass map). 
The slopes of the stellar mass map and the HST/F190N image are basically identical ($-0.044$ and $-0.045$), while the slope of the 3.6\,$\mu m$ image differs significantly ($-0.032$).
The vertical offset determines the $M/L_{1.9\mu m}$ necessary to match both profiles, $M/L_{1.9\mu m} = 0.2\,M_\odot/L_\odot$ (a factor of 3 lower than $M/L_{3.6\mu m}=0.6\,M_\odot/L_\odot$); equivalently, this can be obtained as the ratio of the magnitude zero points, since the HST/F190N and ICA-corrected $M/L_{3.6\mu m}$ profiles expressed in magnitudes clearly overlap (ZP$_{3.6\mu m} = 280.9$\,Jy, ZP$_\mathrm{F190N} = 835.6$\,Jy, leading also to a factor 3 difference).}

\refereetwo{Figure\,\ref{fig:F190N} (left) highlights a clear excess of emission in the centre, especially large in the case of the HST/F190N band, with respect to a simple S\'ersic profile (which shows as a straight line, given that the vertical axis is logarithmic). This excess could be associated with a young stellar population, or it might be contamination from the AGN. The fact that we find $[3.6]-[4.5]>0$ for this region rules out the possibility of a massive central `bulge' or other old inner components. 
In either case, this will bring the results closer to what we measure with the stellar mass map because for young stars the $M/L$ would be considerably lower in the central region and, for an AGN, the excess emission should be ignored.}


\refereetwo{We recomputed the torques and inflow rates assuming that all emission seen at 1.9\,$\mu m$ arisrd from an old stellar population (which would provide a very conservative upper limit, as just argued), and we present the main plots in Fig.\,\ref{fig:F190N} (right). The agreement between the two profiles is very good, except in the region where the HST image displays the `bump'. In principle, the higher spatial resolution afforded by HST ($0.2''$) should allow us to confirm whether inflow persists down to $R = 1''$ (limited by the resolution of PAWS), instead of $R = 1.7''$ (current resolution limit with \textit{Spitzer}). However, as we have emphasised in Sect.\,\ref{Sec:finaluncert}, we would be approaching the dangerous regime where uncertainties become extremely large. Therefore, we highlight again the point that we made before, and consider the inflow measurements highly uncertain in the very central region ($R \lesssim 5''$).}


\section{Hydrodynamical simulations of gas inflow in M51}
\label{sec:appendixsimu}

\citet{2010MNRAS.403..625D} performed N--body and gas simulations of the interaction of M51 and its companion NGC\,5195 and demonstrated that the interaction leads to spiral structure that is remarkably similar to the actual M51. Here we repeat a similar calculation to examine the radial variation of H$_2$ during the encounter, and whether the model predicts a molecular inflow comparable to that observed.  

We show here a simulation similar to that shown in \citet{2010MNRAS.403..625D}, but with some important differences. We used the same orbit for the two galaxies, taken originally from \citet{Theis2003}, and again fully modelled the gas, stars, and dark matter halo using the sphNG code. We used one million particles for the gas, 100,000 particles each for the halo and stellar disc, and 40,000 for the bulge. However, we included a much more massive gas disc (with total gas mass $\sim5\times10^9$ M$_{\odot}$) to better represent the high surface densities in M51. We did not include gas self-gravity, however, so the gas does not undertake gravitational collapse, and the simulation is much easier to run. This also means that we were able to fully model the cold phase of the ISM, including H$_2$ chemistry, effectively. Self-gravity was of course included for the stellar and dark matter components of the simulations. Also, unlike \citet{2010MNRAS.403..625D}, the simulation here is not isothermal, but includes cooling down to a temperature of 10 K and heating (following \citealt{Glover2007} and \citealt{Dobbs2008}), so that the molecular cold HI and warm HI phases of the ISM are modelled. The formation of H$_2$ \citep{Dobbs2008} was also included. The final difference between the simulation here and that of \citet{2010MNRAS.403..625D} is that the companion galaxy was modelled fully with smooth particle hydrodynamics and not with a point mass. However, the companion galaxy is not well resolved, with only 150,000 particles in total. Modelling the companion galaxy has little effect, apart from the fact that the time of the simulation that best matches the current day is slightly earlier. For comparison we also ran a simulation of an isolated galaxy with the same initial conditions as our M51 galaxy.
\begin{figure}
\centerline{\includegraphics[trim=0 60 50 40,clip, scale=0.35]{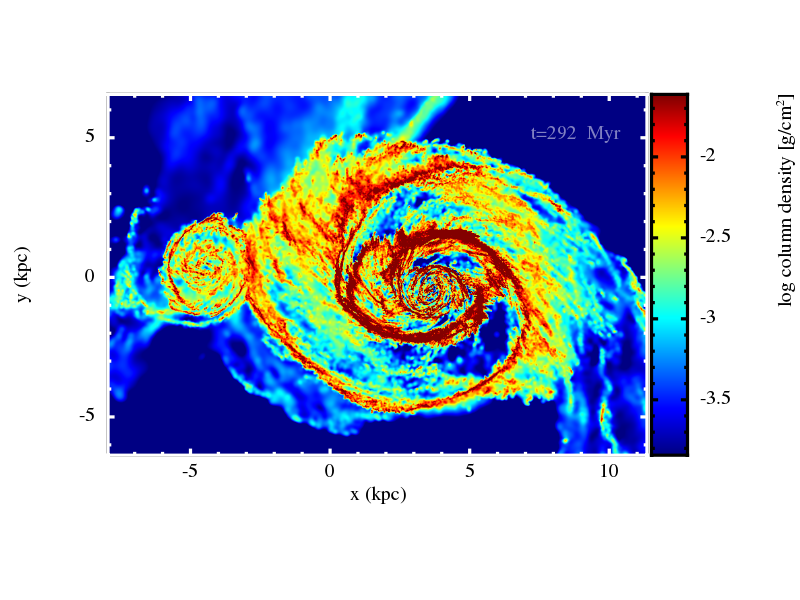}}
\caption{Column density for a simulation of the M51 galaxy interacting with its companion (NGC\,5195). The colour bar is in logarithmic scale, with units of g~cm$^{-2}$. The time shown is 292 Myr, corresponding roughly to the present day.}\label{fig:M51simulation}
\end{figure}

\begin{figure}
\centerline{\includegraphics[scale=0.5]{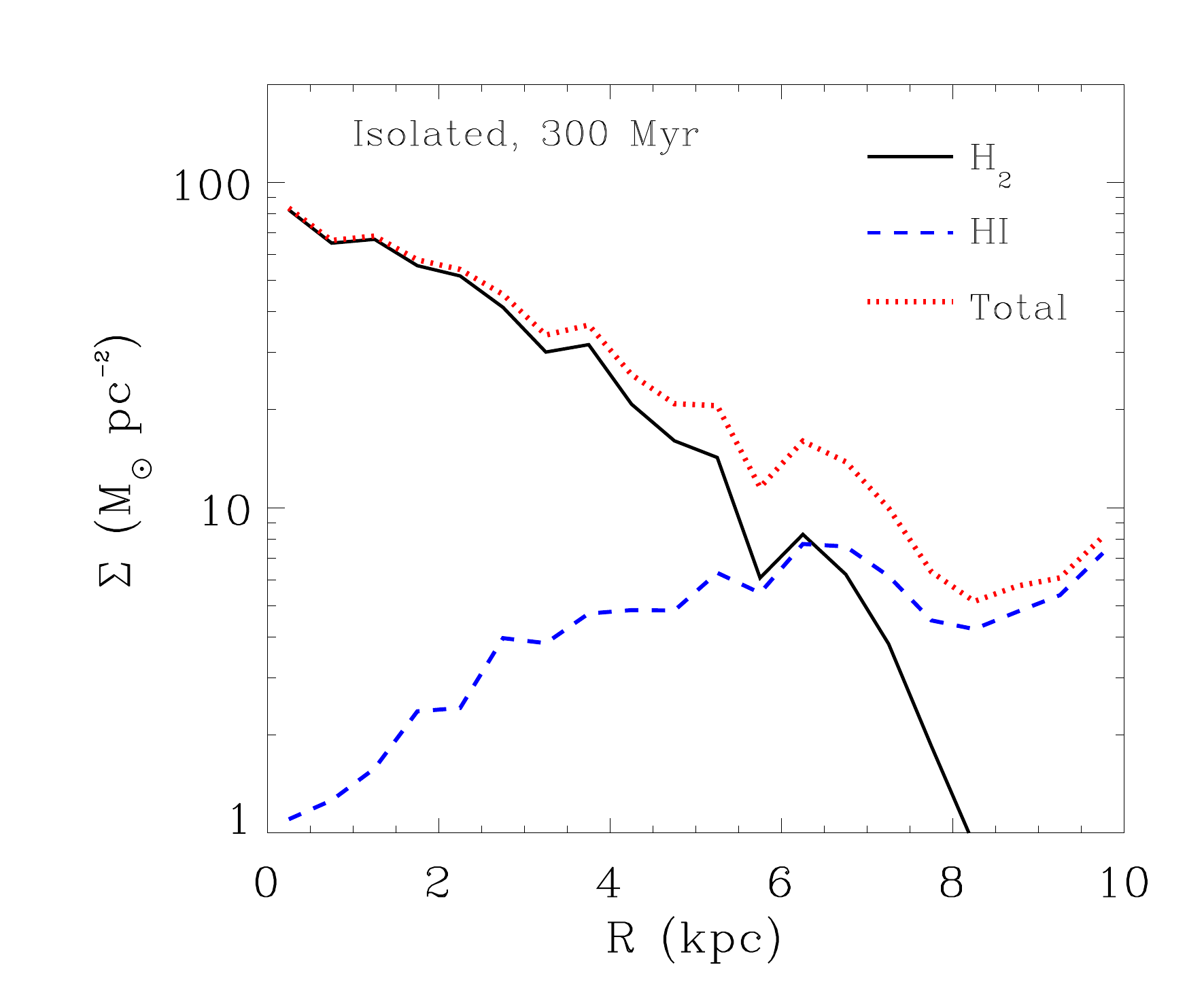}}
\centerline{\includegraphics[scale=0.5]{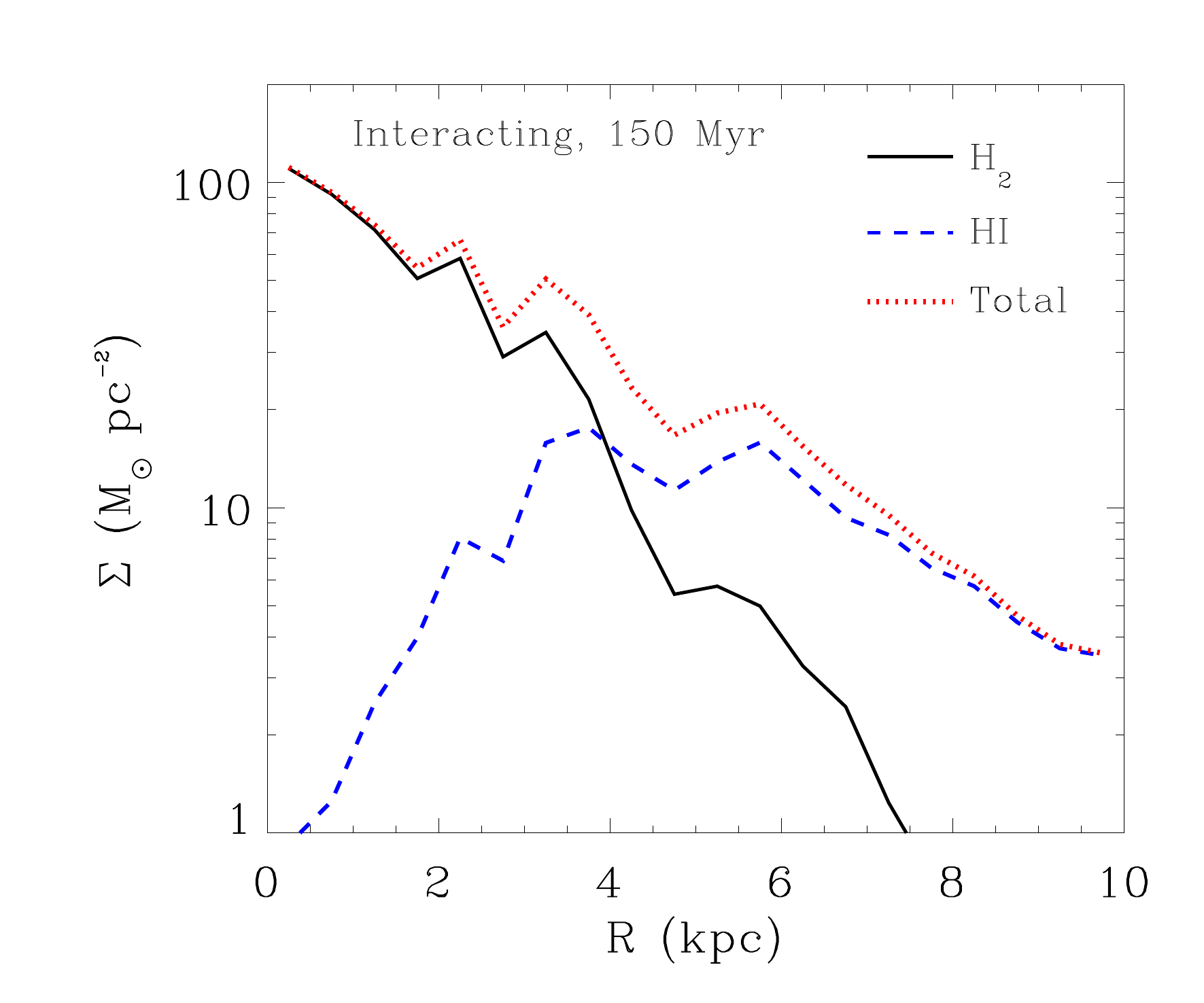}}
\centerline{\includegraphics[scale=0.5]{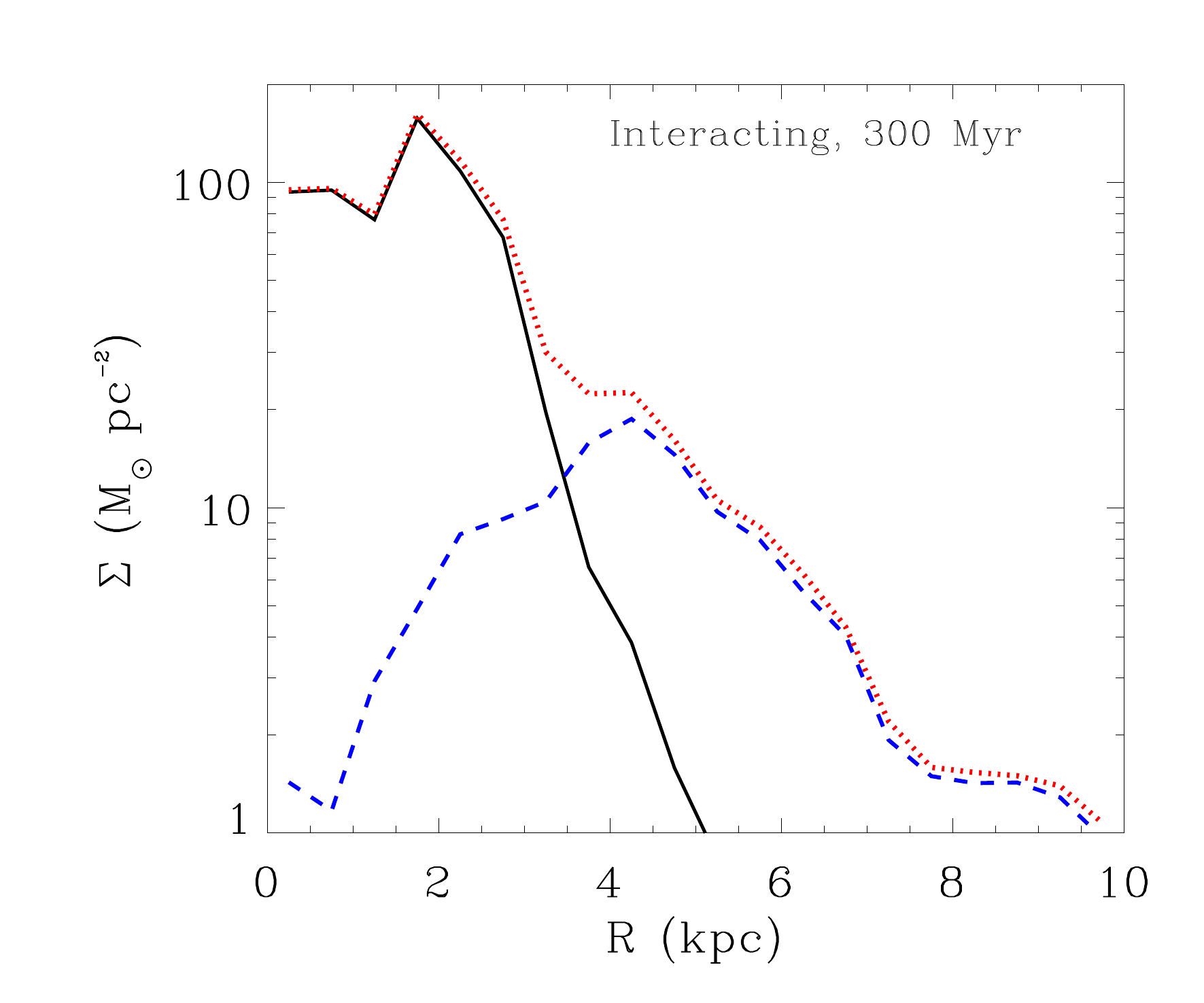}}
\caption{Radial profiles are shown for the molecular, atomic, and total gas surface densities. The top panel shows the profiles for an isolated galaxy example (set up otherwise in the same
way as our model M51 galaxy), and the middle and lower panels show the profiles when our M51 galaxy interacts with a companion.}\label{fig:radialprofiles}
\end{figure}

In Figure~\ref{fig:M51simulation} we show the structure of the M51 galaxy at a time of 292 Myr, corresponding roughly to the present day. The companion is also shown in the figure. The structure is fairly similar to that in \citet{2010MNRAS.403..625D}. The gas shows slightly more structure because it can cool to lower values. Figure~\ref{fig:radialprofiles} shows radial profiles of the atomic, molecular, and total gas for the isolated example (top panel) and the interacting galaxy (middle and lower panels). For the isolated galaxy, the profiles are relatively similar after the first 80 Myr of evolution until the end of the simulation.  The disc is predominantly molecular within 6 kpc, and atomic outside 6 kpc. For the first 150 Myr, the radial profiles of the interacting galaxy evolve similarly to the isolated case, although the molecular gas tends to always be more concentrated. For the time between 150 and 300 Myr, the molecular radial profile is notably more concentrated, with molecular surface densities in excess of 100 M$_{\odot}$ pc$^{-2}$ in the central region (where all the gas is molecular). The molecular surface density profile is consistent with observations by \citet{Leroy2008}.

The interaction of the simulated M51 clearly makes the gas, in particular the molecular gas, more concentrated, which means
that effectively there is inflow towards the centre. We can compare the increase in gas in the centre with the observed inflow rates by making a rough estimate of the change in gas mass in the central 2 kpc. The change in gas mass for a given radial ring is
\begin{equation}
\mathrm{inflow} = \frac{(\hat{\Sigma}_2-\hat{\Sigma}_1)  A} {(T_2-T_1)},
\end{equation}
where $\hat{\Sigma}_2$ is the mean surface density at $T_2$, $\hat{\Sigma}_1$ is the mean surface density at $T_1$, and $A$ is the area of the ring. Here we consider that the area is that of a circle of radius 2 kpc. The timescale for the interaction is somewhat difficult to judge, but the profiles show evident deviations between the isolated and interacting galaxies from 150 to 300 Myr, therefore we took $T_2=300$ Myr and $T_1=150$ Myr. This yields an inflow rate for H$_2$ of  $2.1\pm0.5$ M$_{\odot}$~yr$^{-1}$. This value agrees very well with the observed inflow of M51, suggesting that the interaction of M51 might be responsible for the observed CO inflow.

\bibliography{/Users/querejeta/Documents/E1_SCIENCE_UTIL/mq.bib}{}

\begin{thebibliography}{99}
\expandafter\ifx\csname natexlab\endcsname\relax\def\natexlab#1{#1}\fi

\bibitem[{{Antonucci}(1993)}]{1993ARA&A..31..473A}
{Antonucci}, R. 1993, \araa, 31, 473

\bibitem[{{Athanassoula}(1992{\natexlab{a}})}]{1992MNRAS.259..328A}
{Athanassoula}, E. 1992{\natexlab{a}}, \mnras, 259, 328

\bibitem[{{Athanassoula}(1992{\natexlab{b}})}]{1992MNRAS.259..345A}
{Athanassoula}, E. 1992{\natexlab{b}}, \mnras, 259, 345

\bibitem[{{Barnaby} \& {Thronson}(1992)}]{1992AJ....103...41B}
{Barnaby}, D. \& {Thronson}, Jr., H.~A. 1992, \aj, 103, 41

\bibitem[{{Binney} \& {Tremaine}(1987)}]{1987gady.book.....B}
{Binney}, J. \& {Tremaine}, S. 1987, {Galactic dynamics}

\bibitem[{{Boone} {et~al.}(2007){Boone}, {Baker}, {Schinnerer}, {Combes},
  {Garc{\'{\i}}a-Burillo}, {Neri}, {Hunt}, {L{\'e}on}, {Krips}, {Tacconi}, \&
  {Eckart}}]{2007A&A...471..113B}
{Boone}, F., {Baker}, A.~J., {Schinnerer}, E., {et~al.} 2007, \aap, 471, 113

\bibitem[{{Bradley} {et~al.}(2004){Bradley}, {Kaiser}, \&
  {Baan}}]{2004ApJ...603..463B}
{Bradley}, L.~D., {Kaiser}, M.~E., \& {Baan}, W.~A. 2004, \apj, 603, 463

\bibitem[{{Bresolin} {et~al.}(2004){Bresolin}, {Garnett}, \&
  {Kennicutt}}]{2004ApJ...615..228B}
{Bresolin}, F., {Garnett}, D.~R., \& {Kennicutt}, Jr., R.~C. 2004, \apj, 615,
  228

\bibitem[{{Buta} \& {Combes}(1996)}]{1996FCPh...17...95B}
{Buta}, R. \& {Combes}, F. 1996, Fundamentals of Cosmic Physics, 17, 95

\bibitem[{{Canzian}(1993)}]{1993ApJ...414..487C}
{Canzian}, B. 1993, \apj, 414, 487

\bibitem[{{Ciardullo} {et~al.}(2002){Ciardullo}, {Feldmeier}, {Jacoby}, {Kuzio
  de Naray}, {Laychak}, \& {Durrell}}]{2002ApJ...577...31C}
{Ciardullo}, R., {Feldmeier}, J.~J., {Jacoby}, G.~H., {et~al.} 2002, \apj, 577,
  31

\bibitem[{{Cisternas} {et~al.}(2013){Cisternas}, {Gadotti}, {Knapen}, {Kim},
  {D{\'{\i}}az-Garc{\'{\i}}a}, {Laurikainen}, {Salo},
  {Gonz{\'a}lez-Mart{\'{\i}}n}, {Ho}, {Elmegreen}, {Zaritsky}, {Sheth},
  {Athanassoula}, {Bosma}, {Comer{\'o}n}, {Erroz-Ferrer}, {Gil de Paz}, {Hinz},
  {Holwerda}, {Laine}, {Meidt}, {Men{\'e}ndez-Delmestre}, {Mizusawa},
  {Mu{\~n}oz-Mateos}, {Regan}, \& {Seibert}}]{2013ApJ...776...50C}
{Cisternas}, M., {Gadotti}, D.~A., {Knapen}, J.~H., {et~al.} 2013, \apj, 776,
  50

\bibitem[{{Colombo} {et~al.}(2014){Colombo}, {Meidt}, {Schinnerer},
  {Garc{\'{\i}}a-Burillo}, {Hughes}, {Pety}, {Leroy}, {Dobbs}, {Dumas},
  {Thompson}, {Schuster}, \& {Kramer}}]{2014ApJ...784....4C}
{Colombo}, D., {Meidt}, S.~E., {Schinnerer}, E., {et~al.} 2014, \apj, 784, 4

\bibitem[{{Combes}(2001)}]{2001sac..conf..223C}
{Combes}, F. 2001, in Advanced Lectures on the Starburst-AGN, ed.
  I.~{Aretxaga}, D.~{Kunth}, \& R.~{M{\'u}jica}, 223

\bibitem[{{Combes}(2002)}]{2002astro.ph..8113C}
{Combes}, F. 2002, ArXiv Astrophysics e-prints

\bibitem[{{Combes}(2003)}]{2003ASPC..290..411C}
{Combes}, F. 2003, in Astronomical Society of the Pacific Conference Series,
  Vol. 290, Active Galactic Nuclei: From Central Engine to Host Galaxy, ed.
  S.~{Collin}, F.~{Combes}, \& I.~{Shlosman}, 411

\bibitem[{{Combes} {et~al.}(2004){Combes}, {Garc{\'{\i}}a-Burillo}, {Boone},
  {Hunt}, {Baker}, {Eckart}, {Englmaier}, {Leon}, {Neri}, {Schinnerer}, \&
  {Tacconi}}]{2004A&A...414..857C}
{Combes}, F., {Garc{\'{\i}}a-Burillo}, S., {Boone}, F., {et~al.} 2004, \aap,
  414, 857

\bibitem[{{Combes} {et~al.}(2013){Combes}, {Garc{\'{\i}}a-Burillo}, {Casasola},
  {Hunt}, {Krips}, {Baker}, {Boone}, {Eckart}, {Marquez}, {Neri}, {Schinnerer},
  \& {Tacconi}}]{2013A&A...558A.124C}
{Combes}, F., {Garc{\'{\i}}a-Burillo}, S., {Casasola}, V., {et~al.} 2013, \aap,
  558, A124

\bibitem[{{Combes} {et~al.}(2014){Combes}, {Garc{\'{\i}}a-Burillo}, {Casasola},
  {Hunt}, {Krips}, {Baker}, {Boone}, {Eckart}, {Marquez}, {Neri}, {Schinnerer},
  \& {Tacconi}}]{2014A&A...565A..97C}
{Combes}, F., {Garc{\'{\i}}a-Burillo}, S., {Casasola}, V., {et~al.} 2014, \aap,
  565, A97

\bibitem[{{Combes} \& {Sanders}(1981)}]{1981A&A....96..164C}
{Combes}, F. \& {Sanders}, R.~H. 1981, \aap, 96, 164

\bibitem[{{Comer{\'o}n}(2013)}]{2013A&A...555L...4C}
{Comer{\'o}n}, S. 2013, \aap, 555, L4

\bibitem[{{Comer{\'o}n} {et~al.}(2010){Comer{\'o}n}, {Knapen}, {Beckman},
  {Laurikainen}, {Salo}, {Mart{\'{\i}}nez-Valpuesta}, \&
  {Buta}}]{2010MNRAS.402.2462C}
{Comer{\'o}n}, S., {Knapen}, J.~H., {Beckman}, J.~E., {et~al.} 2010, \mnras,
  402, 2462

\bibitem[{{Comer{\'o}n} {et~al.}(2014){Comer{\'o}n}, {Salo}, {Laurikainen},
  {Knapen}, {Buta}, {Herrera-Endoqui}, {Laine}, {Holwerda}, {Sheth}, {Regan},
  {Hinz}, {Mu{\~n}oz-Mateos}, {Gil de Paz}, {Men{\'e}ndez-Delmestre},
  {Seibert}, {Mizusawa}, {Kim}, {Erroz-Ferrer}, {Gadotti}, {Athanassoula},
  {Bosma}, \& {Ho}}]{2014A&A...562A.121C}
{Comer{\'o}n}, S., {Salo}, H., {Laurikainen}, E., {et~al.} 2014, \aap, 562,
  A121

\bibitem[{{Crane} \& {van der Hulst}(1992)}]{1992AJ....103.1146C}
{Crane}, P.~C. \& {van der Hulst}, J.~M. 1992, \aj, 103, 1146

\bibitem[{{Dobbs} {et~al.}(2008){Dobbs}, {Glover}, {Clark}, \&
  {Klessen}}]{Dobbs2008}
{Dobbs}, C.~L., {Glover}, S.~C.~O., {Clark}, P.~C., \& {Klessen}, R.~S. 2008,
  \mnras, 389, 1097

\bibitem[{{Dobbs} {et~al.}(2010){Dobbs}, {Theis}, {Pringle}, \&
  {Bate}}]{2010MNRAS.403..625D}
{Dobbs}, C.~L., {Theis}, C., {Pringle}, J.~E., \& {Bate}, M.~R. 2010, \mnras,
  403, 625

\bibitem[{{Emsellem} {et~al.}(2015){Emsellem}, {Renaud}, {Bournaud},
  {Elmegreen}, {Combes}, \& {Gabor}}]{2015MNRAS.446.2468E}
{Emsellem}, E., {Renaud}, F., {Bournaud}, F., {et~al.} 2015, \mnras, 446, 2468

\bibitem[{{Ferrarese} \& {Ford}(2005)}]{2005SSRv..116..523F}
{Ferrarese}, L. \& {Ford}, H. 2005, \ssr, 116, 523

\bibitem[{{Ferrarese} \& {Merritt}(2000)}]{2000ApJ...539L...9F}
{Ferrarese}, L. \& {Merritt}, D. 2000, \apjl, 539, L9

\bibitem[{{Font} {et~al.}(2014){Font}, {Beckman}, {Querejeta}, {Epinat},
  {James}, {Blasco-herrera}, {Erroz-Ferrer}, \&
  {P{\'e}rez}}]{2014ApJS..210....2F}
{Font}, J., {Beckman}, J.~E., {Querejeta}, M., {et~al.} 2014, \apjs, 210, 2

\bibitem[{{Fragkoudi} {et~al.}(2015){Fragkoudi}, {Athanassoula}, {Bosma}, \&
  {Iannuzzi}}]{2015MNRAS.450..229F}
{Fragkoudi}, F., {Athanassoula}, E., {Bosma}, A., \& {Iannuzzi}, F. 2015,
  \mnras, 450, 229

\bibitem[{{Garcia-Burillo} {et~al.}(1993){Garcia-Burillo}, {Combes}, \&
  {Gerin}}]{1993A&A...274..148G}
{Garcia-Burillo}, S., {Combes}, F., \& {Gerin}, M. 1993, \aap, 274, 148

\bibitem[{{Garc{\'{\i}}a-Burillo} {et~al.}(2003){Garc{\'{\i}}a-Burillo},
  {Combes}, {Hunt}, {Boone}, {Baker}, {Tacconi}, {Eckart}, {Neri}, {Leon},
  {Schinnerer}, \& {Englmaier}}]{2003A&A...407..485G}
{Garc{\'{\i}}a-Burillo}, S., {Combes}, F., {Hunt}, L.~K., {et~al.} 2003, \aap,
  407, 485

\bibitem[{{Garc{\'{\i}}a-Burillo} {et~al.}(2005){Garc{\'{\i}}a-Burillo},
  {Combes}, {Schinnerer}, {Boone}, \& {Hunt}}]{2005A&A...441.1011G}
{Garc{\'{\i}}a-Burillo}, S., {Combes}, F., {Schinnerer}, E., {Boone}, F., \&
  {Hunt}, L.~K. 2005, \aap, 441, 1011

\bibitem[{{Garc{\'{\i}}a-Burillo} {et~al.}(2014){Garc{\'{\i}}a-Burillo},
  {Combes}, {Usero}, {Aalto}, {Krips}, {Viti}, {Alonso-Herrero}, {Hunt},
  {Schinnerer}, {Baker}, {Boone}, {Casasola}, {Colina}, {Costagliola},
  {Eckart}, {Fuente}, {Henkel}, {Labiano}, {Mart{\'{\i}}n}, {M{\'a}rquez},
  {Muller}, {Planesas}, {Ramos Almeida}, {Spaans}, {Tacconi}, \& {van der
  Werf}}]{2014A&A...567A.125G}
{Garc{\'{\i}}a-Burillo}, S., {Combes}, F., {Usero}, A., {et~al.} 2014, \aap,
  567, A125

\bibitem[{{Garc{\'{\i}}a-Burillo} {et~al.}(2009){Garc{\'{\i}}a-Burillo},
  {Fern{\'a}ndez-Garc{\'{\i}}a}, {Combes}, {Hunt}, {Haan}, {Schinnerer},
  {Boone}, {Krips}, \& {M{\'a}rquez}}]{2009A&A...496...85G}
{Garc{\'{\i}}a-Burillo}, S., {Fern{\'a}ndez-Garc{\'{\i}}a}, S., {Combes}, F.,
  {et~al.} 2009, \aap, 496, 85

\bibitem[{{Gebhardt} {et~al.}(2000){Gebhardt}, {Bender}, {Bower}, {Dressler},
  {Faber}, {Filippenko}, {Green}, {Grillmair}, {Ho}, {Kormendy}, {Lauer},
  {Magorrian}, {Pinkney}, {Richstone}, \& {Tremaine}}]{2000ApJ...539L..13G}
{Gebhardt}, K., {Bender}, R., {Bower}, G., {et~al.} 2000, \apjl, 539, L13

\bibitem[{{Glover} \& {Mac Low}(2007)}]{Glover2007}
{Glover}, S.~C.~O. \& {Mac Low}, M.-M. 2007, \apjs, 169, 239

\bibitem[{{G{\"u}ltekin} {et~al.}(2009){G{\"u}ltekin}, {Richstone}, {Gebhardt},
  {Lauer}, {Tremaine}, {Aller}, {Bender}, {Dressler}, {Faber}, {Filippenko},
  {Green}, {Ho}, {Kormendy}, {Magorrian}, {Pinkney}, \&
  {Siopis}}]{2009ApJ...698..198G}
{G{\"u}ltekin}, K., {Richstone}, D.~O., {Gebhardt}, K., {et~al.} 2009, \apj,
  698, 198

\bibitem[{{Haan} {et~al.}(2009){Haan}, {Schinnerer}, {Emsellem},
  {Garc{\'{\i}}a-Burillo}, {Combes}, {Mundell}, \& {Rix}}]{2009ApJ...692.1623H}
{Haan}, S., {Schinnerer}, E., {Emsellem}, E., {et~al.} 2009, \apj, 692, 1623

\bibitem[{{Heller} \& {Shlosman}(1994)}]{1994ApJ...424...84H}
{Heller}, C.~H. \& {Shlosman}, I. 1994, \apj, 424, 84

\bibitem[{{Ho} {et~al.}(1997{\natexlab{a}}){Ho}, {Filippenko}, \&
  {Sargent}}]{1997ApJS..112..315H}
{Ho}, L.~C., {Filippenko}, A.~V., \& {Sargent}, W.~L.~W. 1997{\natexlab{a}},
  \apjs, 112, 315

\bibitem[{{Ho} {et~al.}(1997{\natexlab{b}}){Ho}, {Filippenko}, \&
  {Sargent}}]{1997ApJ...487..568H}
{Ho}, L.~C., {Filippenko}, A.~V., \& {Sargent}, W.~L.~W. 1997{\natexlab{b}},
  \apj, 487, 568

\bibitem[{{Hohl} \& {Hockney}(1969)}]{1969JCoPh...4..306H}
{Hohl}, F. \& {Hockney}, R.~W. 1969, Journal of Computational Physics, 4, 306

\bibitem[{{Hopkins} {et~al.}(2009){Hopkins}, {Cox}, {Younger}, \&
  {Hernquist}}]{2009ApJ...691.1168H}
{Hopkins}, P.~F., {Cox}, T.~J., {Younger}, J.~D., \& {Hernquist}, L. 2009,
  \apj, 691, 1168

\bibitem[{{Hopkins} \& {Quataert}(2010)}]{2010MNRAS.407.1529H}
{Hopkins}, P.~F. \& {Quataert}, E. 2010, \mnras, 407, 1529

\bibitem[{{Hopkins} \& {Quataert}(2011)}]{2011MNRAS.415.1027H}
{Hopkins}, P.~F. \& {Quataert}, E. 2011, \mnras, 415, 1027

\bibitem[{{Hunt} {et~al.}(2008){Hunt}, {Combes}, {Garc{\'{\i}}a-Burillo},
  {Schinnerer}, {Krips}, {Baker}, {Boone}, {Eckart}, {L{\'e}on}, {Neri}, \&
  {Tacconi}}]{2008A&A...482..133H}
{Hunt}, L.~K., {Combes}, F., {Garc{\'{\i}}a-Burillo}, S., {et~al.} 2008, \aap,
  482, 133

\bibitem[{{Jogee}(2006)}]{2006LNP...693..143J}
{Jogee}, S. 2006, in Lecture Notes in Physics, Berlin Springer Verlag, Vol.
  693, Physics of Active Galactic Nuclei at all Scales, ed. D.~{Alloin}, 143

\bibitem[{{Kennicutt} {et~al.}(2007){Kennicutt}, {Calzetti}, {Walter}, {Helou},
  {Hollenbach}, {Armus}, {Bendo}, {Dale}, {Draine}, {Engelbracht}, {Gordon},
  {Prescott}, {Regan}, {Thornley}, {Bot}, {Brinks}, {de Blok}, {de Mello},
  {Meyer}, {Moustakas}, {Murphy}, {Sheth}, \& {Smith}}]{2007ApJ...671..333K}
{Kennicutt}, Jr., R.~C., {Calzetti}, D., {Walter}, F., {et~al.} 2007, \apj,
  671, 333

\bibitem[{{Knapen} {et~al.}(2000){Knapen}, {Shlosman}, \&
  {Peletier}}]{2000ApJ...529...93K}
{Knapen}, J.~H., {Shlosman}, I., \& {Peletier}, R.~F. 2000, \apj, 529, 93

\bibitem[{{Kormendy} \& {Richstone}(1995)}]{1995ARA&A..33..581K}
{Kormendy}, J. \& {Richstone}, D. 1995, \araa, 33, 581

\bibitem[{{Laine} {et~al.}(2002){Laine}, {Shlosman}, {Knapen}, \&
  {Peletier}}]{2002ApJ...567...97L}
{Laine}, S., {Shlosman}, I., {Knapen}, J.~H., \& {Peletier}, R.~F. 2002, \apj,
  567, 97

\bibitem[{{Leroy} {et~al.}(2009){Leroy}, {Walter}, {Bigiel}, {Usero}, {Weiss},
  {Brinks}, {de Blok}, {Kennicutt}, {Schuster}, {Kramer}, {Wiesemeyer}, \&
  {Roussel}}]{2009AJ....137.4670L}
{Leroy}, A.~K., {Walter}, F., {Bigiel}, F., {et~al.} 2009, \aj, 137, 4670

\bibitem[{{Leroy} {et~al.}(2008){Leroy}, {Walter}, {Brinks}, {Bigiel}, {de
  Blok}, {Madore}, \& {Thornley}}]{Leroy2008}
{Leroy}, A.~K., {Walter}, F., {Brinks}, E., {et~al.} 2008, \aj, 136, 2782

\bibitem[{{Lynden-Bell}(1969)}]{1969Natur.223..690L}
{Lynden-Bell}, D. 1969, \nat, 223, 690

\bibitem[{{Lynden-Bell} \& {Kalnajs}(1972)}]{1972MNRAS.157....1L}
{Lynden-Bell}, D. \& {Kalnajs}, A.~J. 1972, \mnras, 157, 1

\bibitem[{{MacArthur} {et~al.}(2004){MacArthur}, {Courteau}, {Bell}, \&
  {Holtzman}}]{2004ApJS..152..175M}
{MacArthur}, L.~A., {Courteau}, S., {Bell}, E., \& {Holtzman}, J.~A. 2004,
  \apjs, 152, 175

\bibitem[{{Magorrian} {et~al.}(1998){Magorrian}, {Tremaine}, {Richstone},
  {Bender}, {Bower}, {Dressler}, {Faber}, {Gebhardt}, {Green}, {Grillmair},
  {Kormendy}, \& {Lauer}}]{1998AJ....115.2285M}
{Magorrian}, J., {Tremaine}, S., {Richstone}, D., {et~al.} 1998, \aj, 115, 2285

\bibitem[{{Matsushita} {et~al.}(2007){Matsushita}, {Muller}, \&
  {Lim}}]{2007A&A...468L..49M}
{Matsushita}, S., {Muller}, S., \& {Lim}, J. 2007, \aap, 468, L49

\bibitem[{{Matsushita} {et~al.}(2004){Matsushita}, {Sakamoto}, {Kuo}, {Hsieh},
  {Dinh-V-Trung}, {Mao}, {Iono}, {Peck}, {Wiedner}, {Liu}, {Ohashi}, \&
  {Lim}}]{2004ApJ...616L..55M}
{Matsushita}, S., {Sakamoto}, K., {Kuo}, C.-Y., {et~al.} 2004, \apjl, 616, L55

\bibitem[{{Matsushita} {et~al.}(2015){Matsushita}, {Trung}, {Boone}, {Krips},
  {Lim}, \& {Muller}}]{2015ApJ...799...26M}
{Matsushita}, S., {Trung}, D.-V., {Boone}, F., {et~al.} 2015, \apj, 799, 26

\bibitem[{{McConnell} \& {Ma}(2013)}]{2013ApJ...764..184M}
{McConnell}, N.~J. \& {Ma}, C.-P. 2013, \apj, 764, 184

\bibitem[{{Meidt} {et~al.}(2013){Meidt}, {Schinnerer}, {Garc{\'{\i}}a-Burillo},
  {Hughes}, {Colombo}, {Pety}, {Dobbs}, {Schuster}, {Kramer}, {Leroy}, {Dumas},
  \& {Thompson}}]{2013ApJ...779...45M}
{Meidt}, S.~E., {Schinnerer}, E., {Garc{\'{\i}}a-Burillo}, S., {et~al.} 2013,
  \apj, 779, 45

\bibitem[{{Meidt} {et~al.}(2012){Meidt}, {Schinnerer}, {Knapen}, {Bosma},
  {Athanassoula}, {Sheth}, {Buta}, {Zaritsky}, {Laurikainen}, {Elmegreen},
  {Elmegreen}, {Gadotti}, {Salo}, {Regan}, {Ho}, {Madore}, {Hinz}, {Skibba},
  {Gil de Paz}, {Mu{\~n}oz-Mateos}, {Men{\'e}ndez-Delmestre}, {Seibert}, {Kim},
  {Mizusawa}, {Laine}, \& {Comer{\'o}n}}]{2012ApJ...744...17M}
{Meidt}, S.~E., {Schinnerer}, E., {Knapen}, J.~H., {et~al.} 2012, \apj, 744, 17

\bibitem[{{Meidt} {et~al.}(2014){Meidt}, {Schinnerer}, {van de Ven},
  {Zaritsky}, {Peletier}, {Knapen}, {Sheth}, {Regan}, {Querejeta},
  {Mu{\~n}oz-Mateos}, {Kim}, {Hinz}, {Gil de Paz}, {Athanassoula}, {Bosma},
  {Buta}, {Cisternas}, {Ho}, {Holwerda}, {Skibba}, {Laurikainen}, {Salo},
  {Gadotti}, {Laine}, {Erroz-Ferrer}, {Comer{\'o}n}, {Men{\'e}ndez-Delmestre},
  {Seibert}, \& {Mizusawa}}]{2014ApJ...788..144M}
{Meidt}, S.~E., {Schinnerer}, E., {van de Ven}, G., {et~al.} 2014, \apj, 788,
  144

\bibitem[{{Men{\'e}ndez-Delmestre} {et~al.}(2007){Men{\'e}ndez-Delmestre},
  {Sheth}, {Schinnerer}, {Jarrett}, \& {Scoville}}]{2007ApJ...657..790M}
{Men{\'e}ndez-Delmestre}, K., {Sheth}, K., {Schinnerer}, E., {Jarrett}, T.~H.,
  \& {Scoville}, N.~Z. 2007, \apj, 657, 790

\bibitem[{{Mu{\~n}oz-Mateos} {et~al.}(2015){Mu{\~n}oz-Mateos}, {Sheth},
  {Regan}, {Kim}, {Laine}, {Erroz-Ferrer}, {Gil de Paz}, {Comeron}, {Hinz},
  {Laurikainen}, {Salo}, {Athanassoula}, {Bosma}, {Bouquin}, {Schinnerer},
  {Ho}, {Zaritsky}, {Gadotti}, {Madore}, {Holwerda}, {Men{\'e}ndez-Delmestre},
  {Knapen}, {Meidt}, {Querejeta}, {Mizusawa}, {Seibert}, {Laine}, \&
  {Courtois}}]{2015ApJS..219....3M}
{Mu{\~n}oz-Mateos}, J.~C., {Sheth}, K., {Regan}, M., {et~al.} 2015, \apjs, 219,
  3

\bibitem[{{Mundell} \& {Shone}(1999)}]{1999MNRAS.304..475M}
{Mundell}, C.~G. \& {Shone}, D.~L. 1999, \mnras, 304, 475

\bibitem[{{Norris} {et~al.}(2014){Norris}, {Meidt}, {Van de Ven}, {Schinnerer},
  {Groves}, \& {Querejeta}}]{2014ApJ...797...55N}
{Norris}, M.~A., {Meidt}, S., {Van de Ven}, G., {et~al.} 2014, \apj, 797, 55

\bibitem[{{Peng} {et~al.}(2010){Peng}, {Ho}, {Impey}, \&
  {Rix}}]{2010AJ....139.2097P}
{Peng}, C.~Y., {Ho}, L.~C., {Impey}, C.~D., \& {Rix}, H.-W. 2010, \aj, 139,
  2097

\bibitem[{{Pety} {et~al.}(2013){Pety}, {Schinnerer}, {Leroy}, {Hughes},
  {Meidt}, {Colombo}, {Dumas}, {Garc{\'{\i}}a-Burillo}, {Schuster}, {Kramer},
  {Dobbs}, \& {Thompson}}]{2013ApJ...779...43P}
{Pety}, J., {Schinnerer}, E., {Leroy}, A.~K., {et~al.} 2013, \apj, 779, 43

\bibitem[{{Querejeta} {et~al.}(2015){Querejeta}, {Meidt}, {Schinnerer},
  {Cisternas}, {Mu{\~n}oz-Mateos}, {Sheth}, {Knapen}, {van de Ven}, {Norris},
  {Peletier}, {Laurikainen}, {Salo}, {Holwerda}, {Athanassoula}, {Bosma},
  {Groves}, {Ho}, {Gadotti}, {Zaritsky}, {Regan}, {Hinz}, {Gil de Paz},
  {Menendez-Delmestre}, {Seibert}, {Mizusawa}, {Kim}, {Erroz-Ferrer}, {Laine},
  \& {Comer{\'o}n}}]{2015ApJS..219....5Q}
{Querejeta}, M., {Meidt}, S.~E., {Schinnerer}, E., {et~al.} 2015, \apjs, 219, 5

\bibitem[{{Quillen} {et~al.}(1994){Quillen}, {Frogel}, \&
  {Gonzalez}}]{1994ApJ...437..162Q}
{Quillen}, A.~C., {Frogel}, J.~A., \& {Gonzalez}, R.~A. 1994, \apj, 437, 162

\bibitem[{{Quillen} {et~al.}(1995){Quillen}, {Frogel}, {Kenney}, {Pogge}, \&
  {Depoy}}]{1995ApJ...441..549Q}
{Quillen}, A.~C., {Frogel}, J.~A., {Kenney}, J.~D.~P., {Pogge}, R.~W., \&
  {Depoy}, D.~L. 1995, \apj, 441, 549

\bibitem[{{Rautiainen} {et~al.}(2008){Rautiainen}, {Salo}, \&
  {Laurikainen}}]{2008MNRAS.388.1803R}
{Rautiainen}, P., {Salo}, H., \& {Laurikainen}, E. 2008, \mnras, 388, 1803

\bibitem[{{Rix} \& {Rieke}(1993)}]{1993ApJ...418..123R}
{Rix}, H.-W. \& {Rieke}, M.~J. 1993, \apj, 418, 123

\bibitem[{{Salo} {et~al.}(2015){Salo}, {Laurikainen}, {Laine}, {Comer{\'o}n},
  {Gadotti}, {Buta}, {Sheth}, {Zaritsky}, {Ho}, {Knapen}, {Athanassoula},
  {Bosma}, {Laine}, {Cisternas}, {Kim}, {Mu{\~n}oz-Mateos}, {Regan}, {Hinz},
  {Gil de Paz}, {Menendez-Delmestre}, {Mizusawa}, {Erroz-Ferrer}, {Meidt}, \&
  {Querejeta}}]{2015ApJS..219....4S}
{Salo}, H., {Laurikainen}, E., {Laine}, J., {et~al.} 2015, \apjs, 219, 4

\bibitem[{{Schinnerer} {et~al.}(2000){Schinnerer}, {Eckart}, {Tacconi},
  {Genzel}, \& {Downes}}]{2000ApJ...533..850S}
{Schinnerer}, E., {Eckart}, A., {Tacconi}, L.~J., {Genzel}, R., \& {Downes}, D.
  2000, \apj, 533, 850

\bibitem[{{Schinnerer} {et~al.}(2013){Schinnerer}, {Meidt}, {Pety}, {Hughes},
  {Colombo}, {Garc{\'{\i}}a-Burillo}, {Schuster}, {Dumas}, {Dobbs}, {Leroy},
  {Kramer}, {Thompson}, \& {Regan}}]{2013ApJ...779...42S}
{Schinnerer}, E., {Meidt}, S.~E., {Pety}, J., {et~al.} 2013, \apj, 779, 42

\bibitem[{{Schuster} {et~al.}(2007){Schuster}, {Kramer}, {Hitschfeld},
  {Garcia-Burillo}, \& {Mookerjea}}]{2007A&A...461..143S}
{Schuster}, K.~F., {Kramer}, C., {Hitschfeld}, M., {Garcia-Burillo}, S., \&
  {Mookerjea}, B. 2007, \aap, 461, 143

\bibitem[{{Scoville} {et~al.}(2001){Scoville}, {Polletta}, {Ewald}, {Stolovy},
  {Thompson}, \& {Rieke}}]{2001AJ....122.3017S}
{Scoville}, N.~Z., {Polletta}, M., {Ewald}, S., {et~al.} 2001, \aj, 122, 3017

\bibitem[{{Scoville} {et~al.}(1998){Scoville}, {Yun}, {Armus}, \&
  {Ford}}]{1998ApJ...493L..63S}
{Scoville}, N.~Z., {Yun}, M.~S., {Armus}, L., \& {Ford}, H. 1998, \apjl, 493,
  L63

\bibitem[{{Sheth} {et~al.}(2010){Sheth}, {Regan}, {Hinz}, {Gil de Paz},
  {Men{\'e}ndez-Delmestre}, {Mu{\~n}oz-Mateos}, {Seibert}, {Kim},
  {Laurikainen}, {Salo}, {Gadotti}, {Laine}, {Mizusawa}, {Armus},
  {Athanassoula}, {Bosma}, {Buta}, {Capak}, {Jarrett}, {Elmegreen},
  {Elmegreen}, {Knapen}, {Koda}, {Helou}, {Ho}, {Madore}, {Masters},
  {Mobasher}, {Ogle}, {Peng}, {Schinnerer}, {Surace}, {Zaritsky},
  {Comer{\'o}n}, {de Swardt}, {Meidt}, {Kasliwal}, \&
  {Aravena}}]{2010PASP..122.1397S}
{Sheth}, K., {Regan}, M., {Hinz}, J.~L., {et~al.} 2010, \pasp, 122, 1397

\bibitem[{{Shetty} {et~al.}(2007){Shetty}, {Vogel}, {Ostriker}, \&
  {Teuben}}]{2007ApJ...665.1138S}
{Shetty}, R., {Vogel}, S.~N., {Ostriker}, E.~C., \& {Teuben}, P.~J. 2007, \apj,
  665, 1138

\bibitem[{{Shlosman} {et~al.}(1990){Shlosman}, {Begelman}, \&
  {Frank}}]{1990Natur.345..679S}
{Shlosman}, I., {Begelman}, M.~C., \& {Frank}, J. 1990, \nat, 345, 679

\bibitem[{{Shlosman} {et~al.}(1989){Shlosman}, {Frank}, \&
  {Begelman}}]{1989Natur.338...45S}
{Shlosman}, I., {Frank}, J., \& {Begelman}, M.~C. 1989, \nat, 338, 45

\bibitem[{{Shlosman} \& {Noguchi}(1993)}]{1993ApJ...414..474S}
{Shlosman}, I. \& {Noguchi}, M. 1993, \apj, 414, 474

\bibitem[{{Sparke} \& {Sellwood}(1987)}]{1987MNRAS.225..653S}
{Sparke}, L.~S. \& {Sellwood}, J.~A. 1987, \mnras, 225, 653

\bibitem[{{Tagger} {et~al.}(1987){Tagger}, {Sygnet}, {Athanassoula}, \&
  {Pellat}}]{1987ApJ...318L..43T}
{Tagger}, M., {Sygnet}, J.~F., {Athanassoula}, E., \& {Pellat}, R. 1987, \apjl,
  318, L43

\bibitem[{{Theis} \& {Spinneker}(2003)}]{Theis2003}
{Theis}, C. \& {Spinneker}, C. 2003, \apss, 284, 495

\bibitem[{{van der Laan} {et~al.}(2011){van der Laan}, {Schinnerer}, {Boone},
  {Garc{\'{\i}}a-Burillo}, {Combes}, {Haan}, {Leon}, {Hunt}, \&
  {Baker}}]{2011A&A...529A..45V}
{van der Laan}, T.~P.~R., {Schinnerer}, E., {Boone}, F., {et~al.} 2011, \aap,
  529, A45

\bibitem[{{van der Laan} {et~al.}(2013){van der Laan}, {Schinnerer},
  {Emsellem}, {Meidt}, {Dumas}, {B{\"o}ker}, {Hunt}, {Haan}, {Mundell}, \&
  {Wozniak}}]{2013A&A...556A..98V}
{van der Laan}, T.~P.~R., {Schinnerer}, E., {Emsellem}, E., {et~al.} 2013,
  \aap, 556, A98

\bibitem[{{Wada} \& {Koda}(2004)}]{2004MNRAS.349..270W}
{Wada}, K. \& {Koda}, J. 2004, \mnras, 349, 270

\bibitem[{{Wainscoat} {et~al.}(1989){Wainscoat}, {Freeman}, \&
  {Hyland}}]{1989ApJ...337..163W}
{Wainscoat}, R.~J., {Freeman}, K.~C., \& {Hyland}, A.~R. 1989, \apj, 337, 163

\bibitem[{{Walter} {et~al.}(2008){Walter}, {Brinks}, {de Blok}, {Bigiel},
  {Kennicutt}, {Thornley}, \& {Leroy}}]{2008AJ....136.2563W}
{Walter}, F., {Brinks}, E., {de Blok}, W.~J.~G., {et~al.} 2008, \aj, 136, 2563

\bibitem[{{Wong} {et~al.}(2004){Wong}, {Blitz}, \&
  {Bosma}}]{2004ApJ...605..183W}
{Wong}, T., {Blitz}, L., \& {Bosma}, A. 2004, \apj, 605, 183

\bibitem[{{Zhang} \& {Buta}(2012)}]{2012arXiv1203.5334Z}
{Zhang}, X. \& {Buta}, R.~J. 2012, ArXiv e-prints

\bibitem[{{Zibetti} {et~al.}(2009){Zibetti}, {Charlot}, \&
  {Rix}}]{2009MNRAS.400.1181Z}
{Zibetti}, S., {Charlot}, S., \& {Rix}, H.-W. 2009, \mnras, 400, 1181

\end{thebibliography}
\bibliographystyle{aa}{}

\end{document}